\documentclass{cimento}
\usepackage{epsf}
\title{Exotic solutions in string theory}
\shorttitle{Exotic solutions in string theory}
\author{S.V.Klimenko\from{ins:x}, I.N.Nikitin\from{ins:y}}
\instlist{
\inst{ins:x} Institute for High Energy Physics, Protvino, Russia
\inst{ins:y} National Research Center for Information Technology,
St.Augustin, Germany}
\PACSes{
\PACSit{11.25}{theory of fundamental strings}
\PACSit{11.27}{extended classical solutions}
}

\newlength{\mywidth}\mywidth=2.75truein %width of caption
\renewenvironment{figure}{\refstepcounter{figure}
\baselineskip=0.4\normalbaselineskip\footnotesize}
{\baselineskip=\normalbaselineskip}
\def\fignum{{\bf Fig.\arabic{figure}.\quad}}
\def\fref#1{fig.\ref{#1}}

\def\frac#1#2{{\textstyle{{#1}\over{#2}}}}

\def\s{\sigma}
\def\Z{{\bf Z}}

\def\half{{\textstyle{{1}\over{2}}}}
\def\quart{{\textstyle{{1}\over{4}}}}
\def\ran{\rangle}
\def\vv{\Bigr|}
\def\sgn{\mathop{\rm sign}\nolimits}
\def\fproof#1{\footnote{#1}.}
\def\vsp{\vspace{0.3cm}}
\def\nn{\nonumber}
\def\ph{\varphi}
\def\twin#1#2{\scriptstyle #1\atop\scriptstyle #2}
\def\df{\partial}
\def\L{{\cal L}}

\begin{document}
\maketitle

\begin{abstract}
Solutions of classical string theory, correspondent to the world sheets, 
mapped in Minkowsky space with a fold, are considered.
Typical processes for them are creation of strings from vacuum,
their recombination and annihilation. These solutions violate
positiveness of square of mass and Regge condition. In quantum string theory
these solutions correspond to physical states $|DDF\ran+|sp\ran$
with non-zero spurious component.
\end{abstract}

\section*{Introduction}
In this work we want to draw theoretists' attention 
to the fact that classical mechanics of Nambu-Goto string
in its covariant Hamiltonian formulation contains
solutions with quite exotic properties. Particularly,
it contains solutions with {\it negative square of mass}.
One can easily clarify this in the following example.

\vspace{3mm}\noindent\underline{{\it Example 1.}}
Phase space of open string is described by infinite
set of canonical oscillator variables $a_{\mu}^{n}$,
restricted by reality condition $(a_{\mu}^{n})^{*}=a_{\mu}^{-n}$
and Virasoro constraints
$L^{n}=\sum_{k} a_{\mu}^{k}a_{\mu}^{n-k}=0$.
Let consider the following set of oscillators:
$$a^{0}_{\mu}=(1,\alpha,0),\ a^{\pm1}_{\mu}=(\alpha,1/2,\pm1/2i),\ 
a^{\pm2}_{\mu}=(0,\alpha/2,\pm\alpha/2i),\ \mbox{others}~a^{n}_{\mu}=0.$$
Here we consider a theory in 3-dimensional Minkowsky space,
$a_{\mu}^{n}=(a_{0}^{n},a_{1}^{n},a_{2}^{n})$;
$\alpha$ is real parameter.

Let's show, that given set satisfies Virasoro constraints.
At $|n|>4\ L^{n}=0$, because in each term of the sum
$\sum a_{\mu}^{k}a_{\mu}^{n-k}$ one of oscillator variables vanishes.
Due to a property $L^{-n}=(L^{n})^{*}$, the check is needed
only for conditions $L^{n}=0$ at $0\leq n\leq4$. 
These conditions have a form:
$$(a^{2}_{\mu})^{2}=0,\ a^{1}_{\mu}a^{2}_{\mu}=0,\ 
2a^{0}_{\mu}a^{2}_{\mu}+(a^{1}_{\mu})^{2}=0,\
a^{-1}_{\mu}a^{2}_{\mu}+a^{0}_{\mu}a^{1}_{\mu}=0,\ 
(a^{0}_{\mu})^{2}+2a^{-1}_{\mu}a^{1}_{\mu}+2a^{-2}_{\mu}a^{2}_{\mu}=0,$$
their validity can be easily proven by direct substitution of
$a_{\mu}^{n}$.

Total momentum of the string is defined by an expression:
$P_{\mu}=\sqrt{\pi}a^{0}_{\mu}$. Thus, square of string's mass
$P^{2}=\pi(1-\alpha^{2})$ is positive at $|\alpha|<1$ and negative at
$|\alpha|>1$. 

\vspace{3mm}
Let's consider a function 
$a_{\mu}(\s)=1/\sqrt{\pi}\sum_{n} a_{\mu}^{n}e^{in\s}$.
In given example $a_{0}(\s)=(1+2\alpha\cos\s)/\sqrt{\pi}$, at $|\alpha|>1/2$
this function is not everywhere positive. 
Solutions, for which $a_{0}(\s)$ has variable sign, will be further 
called {\it exotic}. In this paper we will consider the properties
of such solutions. In Section~1 we describe the geometrical method
for reconstruction of string dynamics and show
several examples of exotic dynamics.
In Section~2 we consider the properties of exotic 
solutions in Lagrangian theory.
In Section~3 the appearance of exotic solutions in quantum string theory
is discussed.

\section{Exotic solutions in Hamiltonian theory}
\paragraph*{1.1. Geometrical reconstruction of exotic solutions.}\ 

Let's introduce a function, related with
string's coordinates and momenta
by expressions~\cite{zone}

\vspace{-5mm}
\begin{eqnarray}
&&Q_{\mu}(\s)=x_{\mu}(\s)+\int_{0}^{\s}d\tilde\s \;
p_{\mu}(\tilde\s),\label{Qini0}\\
&&x_{\mu}(\s)=(Q_{\mu}(\s)+Q_{\mu}(-\s))/2,\quad
p_{\mu}(\s)=(Q'_{\mu}(\s)+Q'_{\mu}(-\s))/2\label{Qini}
\end{eqnarray}
($x,p$ are {\it even} functions of $\s$).
In terms of oscillator variables, introduced earlier:
\begin{eqnarray}
&&Q_{\mu}(\s)=X_{\mu}+\frac{P_{\mu}}{\pi}\sigma+
{\textstyle{{1}\over{\sqrt{\pi}}}}
\sum\limits_{n\neq0}{\textstyle{{a_{\mu}^{n}}\over{in}}}e^{in\sigma},\quad
Q_{\mu}'(\s)=a_{\mu}(\s).\label{Qosc}
\end{eqnarray}
The curve, defined by the function $Q_{\mu}(\s)$ (further called
{\it supporting curve}) has the following properties:

\vsp
\noindent 1.~the curve is light-like: $Q'^{2}_{\mu}(\s)=0$, 
this property is equivalent to Virasoro constraints 
on oscillator variables;

\noindent 2.~the curve is periodical: 
$Q_{\mu}(\s+2\pi)-Q_{\mu}(\s)=Const=2P_{\mu}$\quad
($P_{\mu}$ is total momentum of the string);

\noindent 3.~the curve coincides with the world line of one string end:
$x_{\mu}(0,\tau)=Q_{\mu}(\tau)$; the world line of another end
is the same curve, shifted onto the semi-period:
$x_{\mu}(\pi,\tau)=Q_{\mu}(\pi+\tau)-P_{\mu}$;

\noindent 4.~the whole world sheet is reconstructed by this curve
as follows: $x_{\mu}(\s,\tau)=(Q_{\mu}(\s_{1})+Q_{\mu}(\s_{2}))/2$, 
$\s_{1,2}=\tau\pm\s$, see \fref{f1}.

\vsp\noindent
These properties can be easily proven from the definition of $Q_{\mu}(\s)$
and known mechanics in oscillator variables, see Appendix~1.

Further consideration will be restricted to the supporting curves,
whose time component $Q_{0}(\s)$ is non-monotonous function,
see \fref{f2}. Such curves can be explicitly constructed, specifying
tangent vector in the form $Q_{\mu}'(\s)=a_{0}(\s)(1,\vec n(\s)),$
where $\vec n(\s)^{2}=1$, $\vec n(\s)$ is $2\pi$-periodic function,
and $a_{0}(\s)$ is $2\pi$-periodic function of variable sign\footnote{
Substitution $\vec n(\s)=(\cos\s,\sin\s)$ and 
$a_{0}(\s)=(1+2\alpha\cos\s)/\sqrt{\pi}$ corresponds to oscillator variables
from Example~1.}. Such supporting curves necessarily have singularities
(cusps)\fproof{When $Q_{0}'(\s)$ changes its sign, vector
$Q_{\mu}'(\s)$, lying on the light cone, 
passes through the origin: $Q_{\mu}'(\s^{*})=0$.
In this point the supporting curve has a cusp.}

\begin{center}
\parbox{7cm}{\begin{figure}\label{f1}
\begin{center}
~\epsfysize=4.5cm\epsfxsize=6cm\epsffile{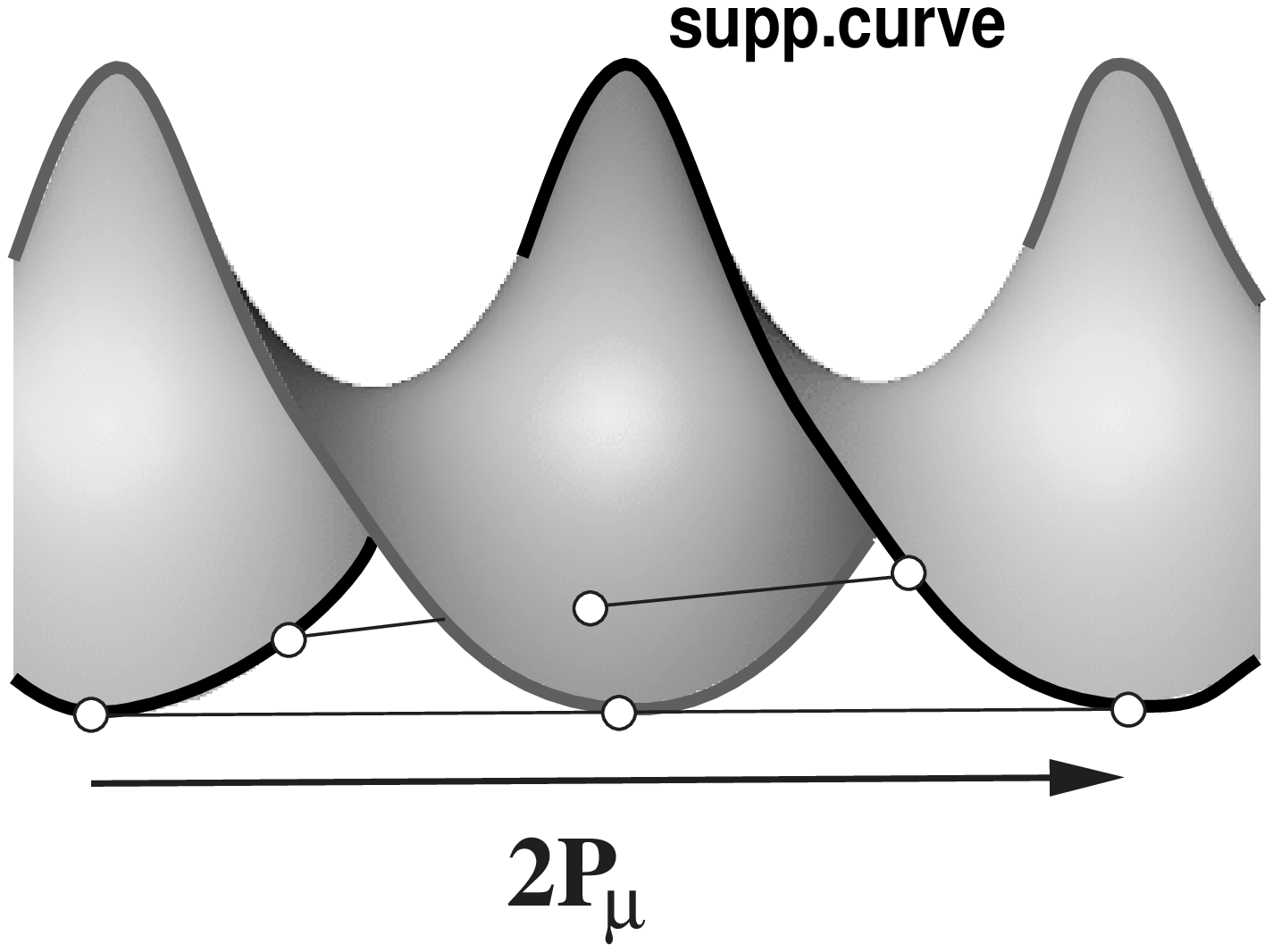}
\end{center}

\fignum World sheet of open string is constructed as
a locus of middles of segments,
connecting all possible pairs of points on the supporting curve.

\end{figure}
}\quad\quad
\parbox{5cm}{\begin{figure}\label{f2}
\begin{center}
~\epsfysize=4.5cm\epsfxsize=4cm\epsffile{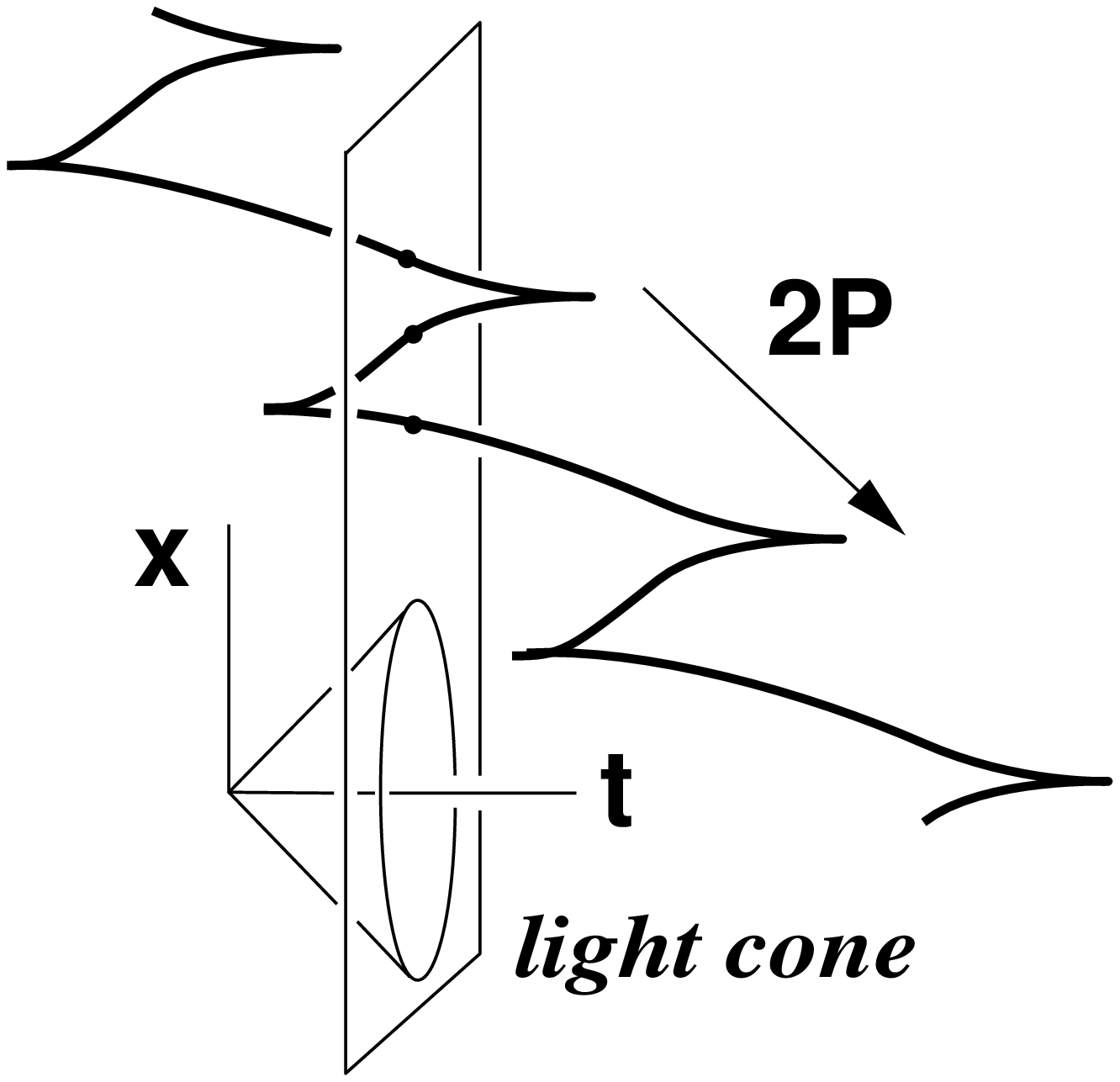}
\end{center}

\fignum Supporting curve, non-monotonous along time axis. 

\end{figure}
}
\end{center}

For the curves with $a_{0}(\s)>0$ (correspondent solutions are further called
{\it normal}) vector $P_{\mu}$ is time-like or light-like: $P^{2}\geq0$,
because $P_{\mu}=\int_{0}^{2\pi} d\s a_{\mu}(\s)/2$ is represented as 
a sum of light-like vectors, directed into the future. 
For non-monotonous supporting curves $P^{2}$ might
take the both signs. At first, let's study the case of time-like
$P_{\mu}$. In this case we can consider the evolution in center-of-mass frame 
(CMF), where $P_{\mu}$ is directed along the time axis.

\vspace{3mm}\noindent\underline{{\it Example 2}}. 
Let each period of supporting curve
contains a single non-monotonous interval, whose sizes are
small comparing with the period. E.g. 
$Q_{\mu}'(\s)=a_{0}(\s)(1,\cos\s,\sin\s)$
with $a_{0}(\s)=1+1.4\cos2\s -0.6\sin3\s$. 
For this choice $P_{\mu}=(\pi,0,0)$
is directed along the time axis. The graph of function $a_{0}(\s)$
and correspondent supporting curve are shown on \fref{f3a}.

Simplified model of the world sheet, stretched onto such supporting curve,
is shown on \fref{f3}. Equal-time slice of the world sheet 
at $t<t_{A}$ is a single connected curve (``permanent component''),
consisting of two segments $L$ and $a$. At $t=t_{A}$
an additional open string $bc$ appears. At $t=t_{R}$
a process of recombination occurs,
the strings exchange by their segments: 
$(La)+(bc)\to(Lb)+(ac)$. Short string $ac$ disappears at $t=t_{B}$.
Further evolution (till the end of a period) contains only 
permanent component $Lb$. 

\begin{figure}\label{f5}
\begin{center}
\parbox{7cm}
{~\epsfysize=9cm\epsfxsize=7cm\epsffile{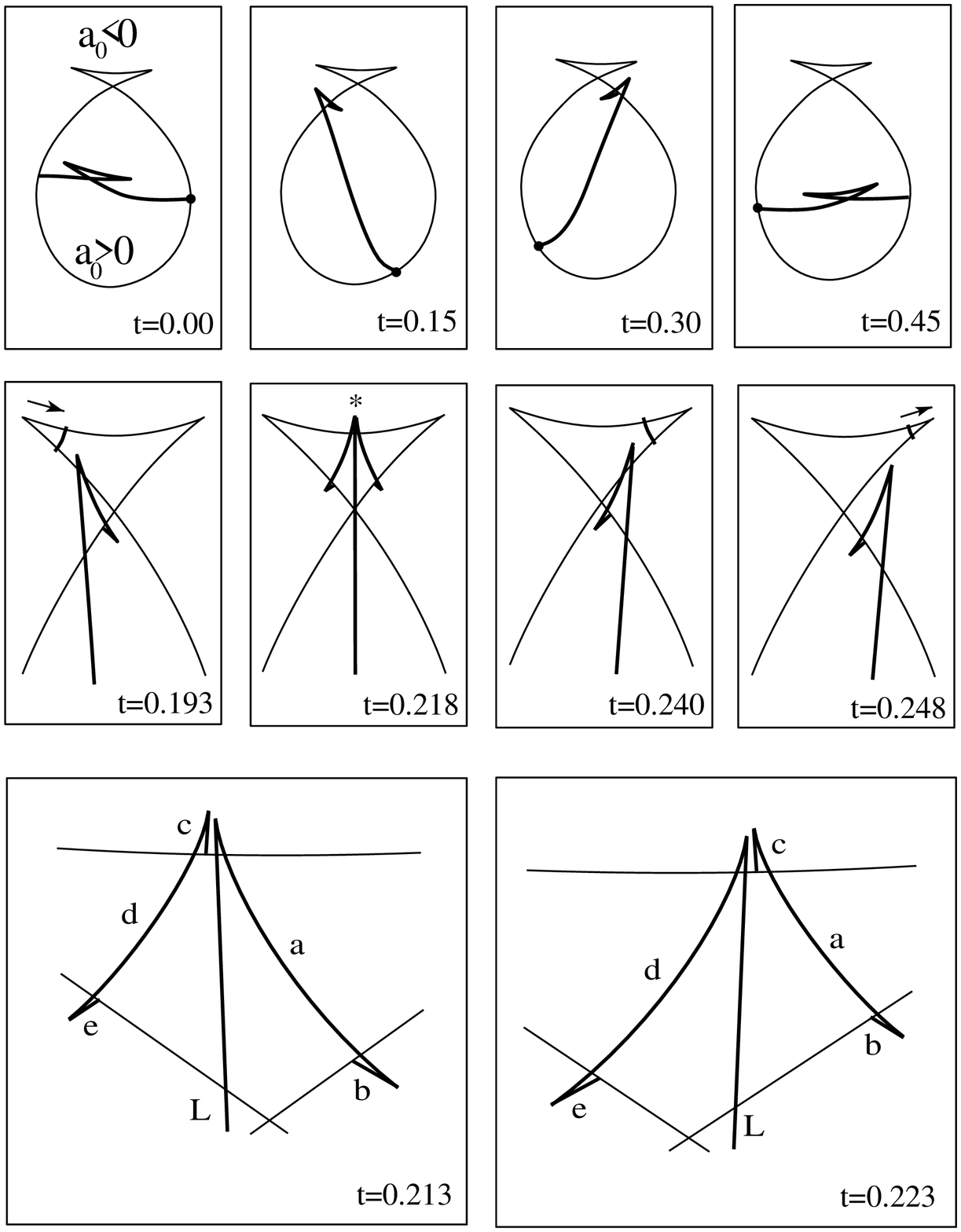}}
\quad\quad\parbox{5cm}
{\fignum
String dynamics for Example 2.
Bold lines are the strings, thin line is the
supporting curve (projected to space component of CMF). The supporting curve
has two segments: $a_{0}>0$ and $a_{0}<0$, directed 
in Minkowsky space forward and backward in time respectively. 
There is a long string, which is permanently present
in the system. The common features of its evolution are shown on the upper
frame sequence. Time is measured in periods. 
Central sequence shows in details the time interval, when the end of the string
passes the part with $a_{0}<0$. Additional short string appears near $t=0.171$.
At $t=0.218$ the recombination ($*$) occurs: 
short string is attached to the long string and a part of long 
string is detached. New short string disappears at $t=0.265$. 
Lower frames show the details of recombination process:
$(Lab)+(cde)\to(cab)+(Lde)$, segments $L,c$ are exchanged.  
}
\end{center}
\end{figure}

Realistic image of the world sheet (\fref{f4}) contains more details.
Particularly, one can find here singular lines
($fRAd$ and $gBRe$). Slices of the world sheet have cusps along these 
lines. These two lines are congruent and homotetic to 
to the supporting curve ($cABh$) with the coefficient $1/2$.
They are created by cusp points $A,B$ on the supporting 
curve\fproof{Cusp on supporting curve in point $Q(\s^{*})$
creates a singularity on the world sheet,
located along the line $(Q(\s^{*})+Q(\s))/2$,
i.e. the supporting curve, contracted twice to the point
$Q(\s^{*})$.}
General structure of equal-time slices is the same:
creation of new open string in point $A$,
its recombination with the permanent component in point $R$,
and annihilation of new string in point $B$.
Equal-time slices of this surface are shown on \fref{f5}.

\begin{figure}\label{f5b}
\begin{center}
\parbox[c]{7cm}{
~\epsfysize=7cm\epsfxsize=7cm\epsffile{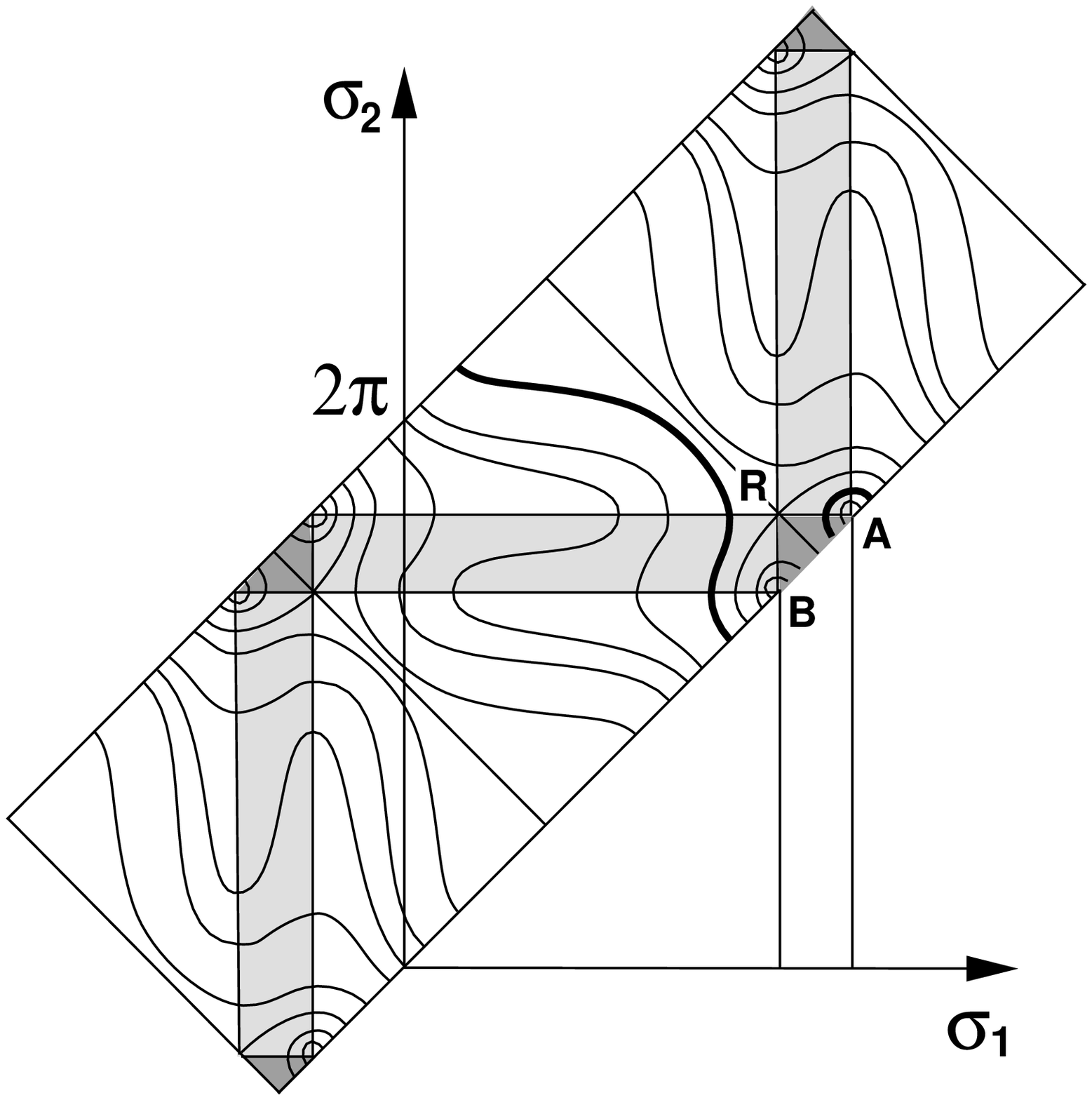}
}\quad
\quad\quad\parbox[c]{5cm}{
\begin{center}
~\epsfysize=4.8cm\epsfxsize=4.8cm\epsffile{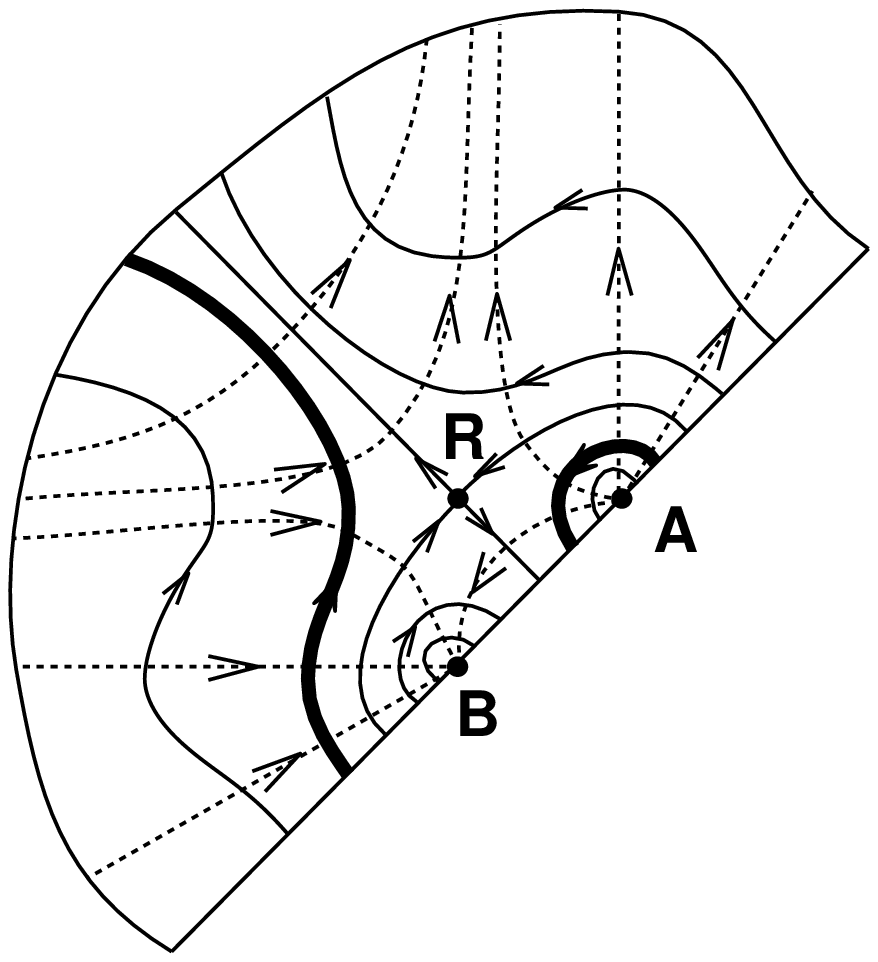}
\end{center}

\fignum Equal-time slices on parameters plane. 
Right part shows details of recombination process. 
Arrows on strings (solid lines) show their orientation.
Arrows on dashed lines show the direction of evolution.
}
\end{center}
\end{figure}

Fig.\ref{f5b} shows equal-time slices $x_{0}(\s_{1},\s_{2})=t$
on the plane of parameters $(\s_{1},\s_{2})$.  
Band $\s_{1}\leq\s_{2}\leq\s_{1}+2\pi$ on this plane
uniquely represents the world sheet.
Slice for the moment $t:\ t_{A}<t<t_{R}$ is shown on this figure
in bold (two disconnected parts). In further evolution small part
is expanded and recombinates with
long part in point $R$. After recombination new small part is shrinked in
point $B$, and new long part continues to move in right-upward direction.

Straight lines, passing through points $\s_{i}=\s_{A,B}+2\pi n,\
i=1,2,\ n\in\Z$, correspond to cusp lines on the world sheet.
On these lines $Q'_{0}(\s)$ changes its sign.
Vertical grey bands show regions with $Q'_{0}(\s_{1})<0$,
horizontal grey bands correspond to $Q'_{0}(\s_{2})<0$.

\begin{figure}\label{f6a}
\begin{center}
~\epsfysize=3cm\epsfxsize=5cm\epsffile{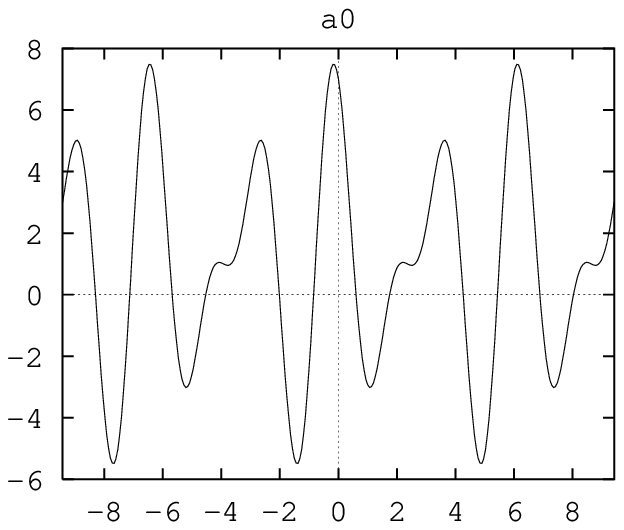}\quad
~\epsfysize=3cm\epsfxsize=7cm\epsffile{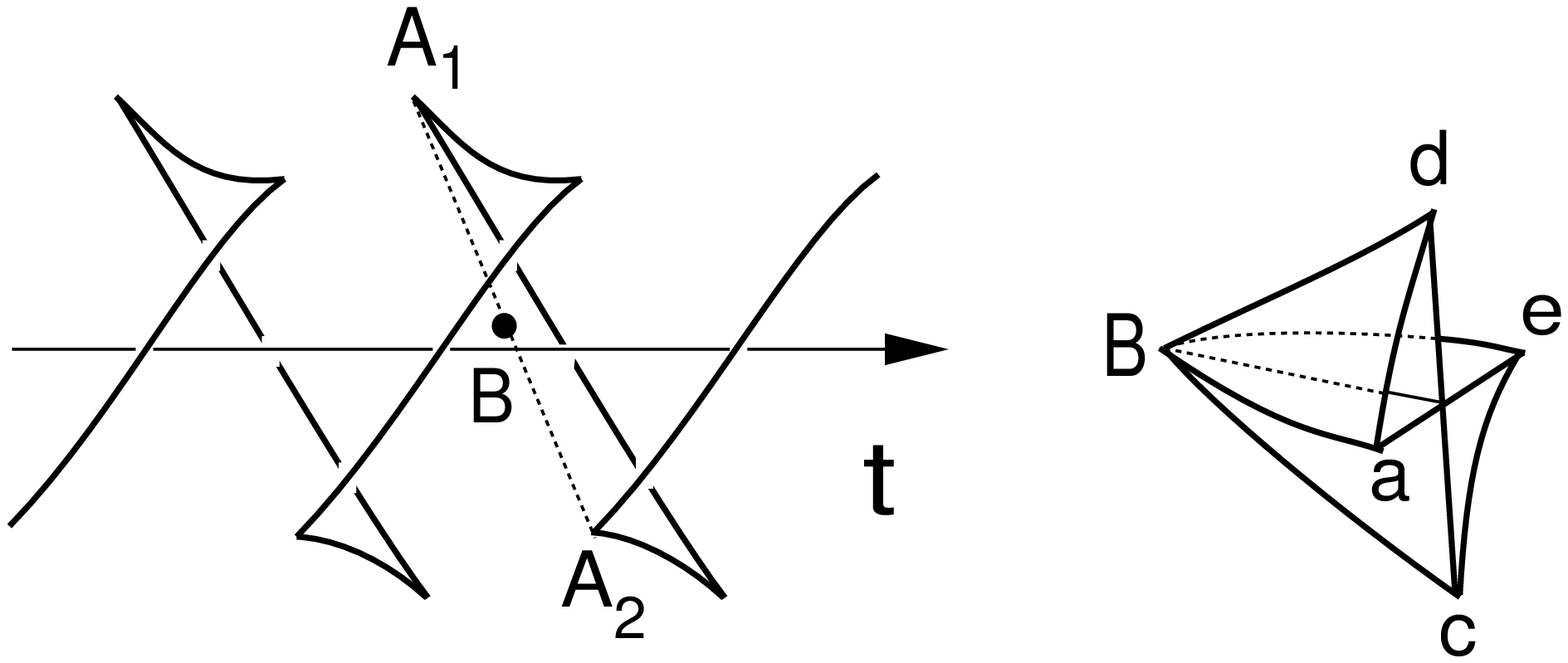}
\end{center}

\noindent
\fignum Function $a_{0}(\s)$ and supporting curve for Example 3.
Two cusps of supporting curve $A_{1}$ and $A_{2}$
create in the vicinity of their middle point $B=(A_{1}+A_{2})/2$
a surface, shown on the right part of the figure. Here the lines
$aBc$ and $dBe$ are homotetic with the coefficient $1/2$ 
to the parts of supporting curves near cusps $A_{1}$ and $A_{2}$
respectively. When the curve $aBc$ moves parallelly to itself,
keeping the vertex on the curve $dBe$, it spans the world sheet. 
The part of the surface
$(aBd+cBd)$ is spanned by the curve $aBc$ in motion 
along segment $Bd$, the part $(aBe+cBe)$ is spanned by $aBc$
in motion along $Be$. Equal time slice of the obtained surface
contains closed string $aecd$.

\end{figure}

\begin{figure}\label{f6}
\begin{center}
~\epsfysize=8cm\epsfxsize=8cm\epsffile{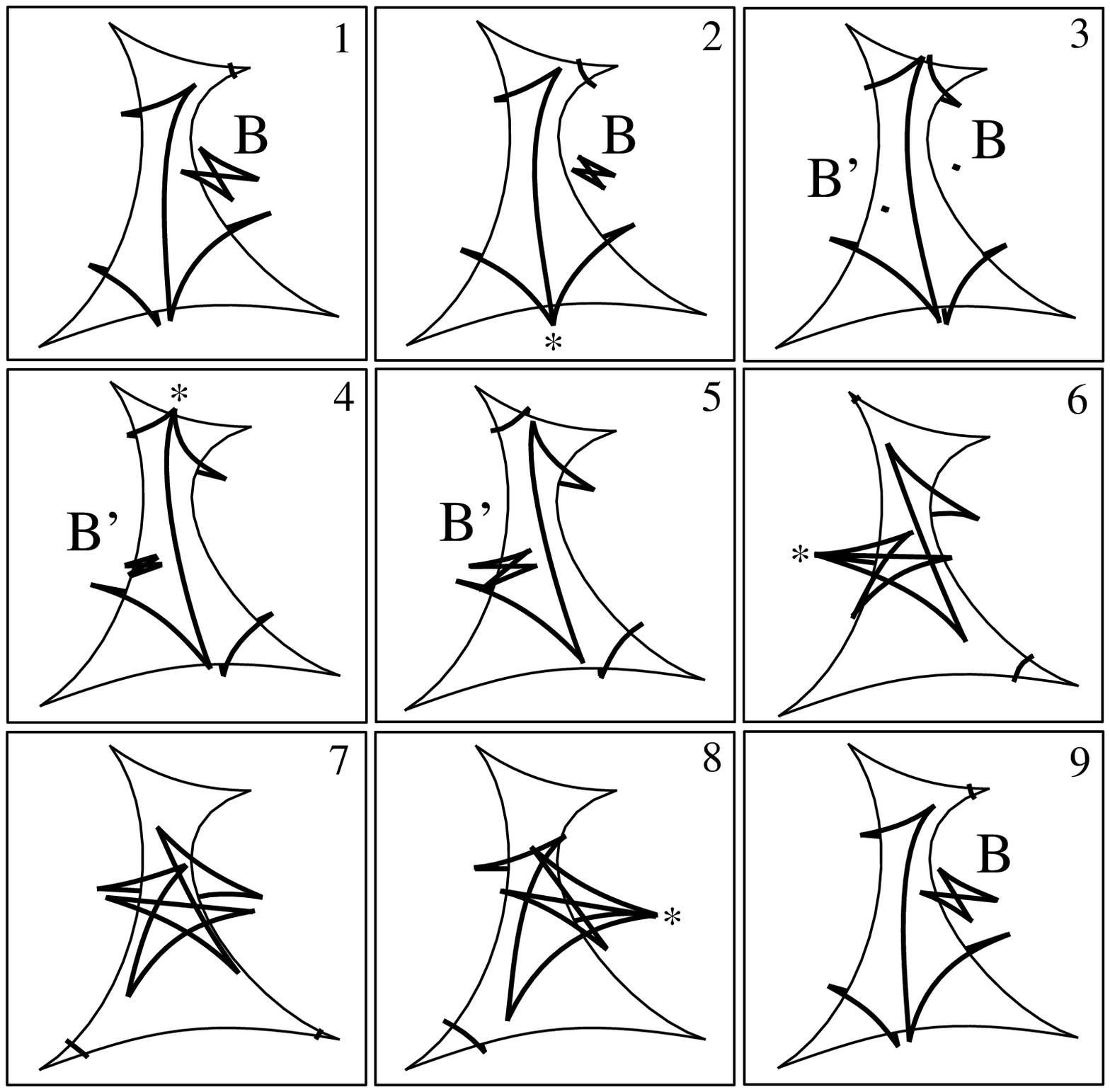}
\end{center}

\noindent\fignum
String dynamics for Example 3. On the first
frame the following 4 disconnected elements can be found: Z-shaped long open
string in the center, two shorter open strings in the corners of 
supporting curve and closed string $B$, looking like a butterfly. 
In the process of further evolution the long
string recombinates with shorter strings, adopting the shape of reverted Z
(frame~5). On frame~3 the closed string $B$ disappears, 
and new closed string $B'$ appears. 
On frame~6 it enters into recombination with the long string and is 
included in it. There are no closed strings on frame~7, 
here the long string in the center is connected.
On the frame~8 the string $B$ is detached from the long string. The 
frame~9 coincides with the frame~1, and the evolution is repeated again.

\end{figure}

\vspace{3mm}\noindent\underline{{\it Example 3.}} 
$a_{0}(\s)=1+4\cos2\s+2\cos3\s-2\sin3\s$, see \fref{f6a}. 
This supporting curve contains two non-monotonous 
intervals in each period. Their interference leads to more 
complicated processes, particularly, appearance of 
closed strings in the evolution, see \fref{f6}.

\begin{center}
\parbox{6cm}{
\begin{figure}\label{f3a}
\begin{center}
~\hspace{-5mm}~\epsfysize=3cm\epsfxsize=6cm\epsffile{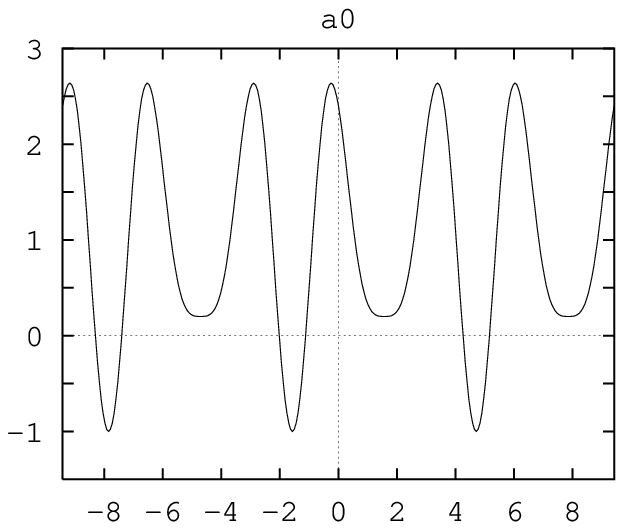}

~\epsfysize=2cm\epsfxsize=5.4cm\epsffile{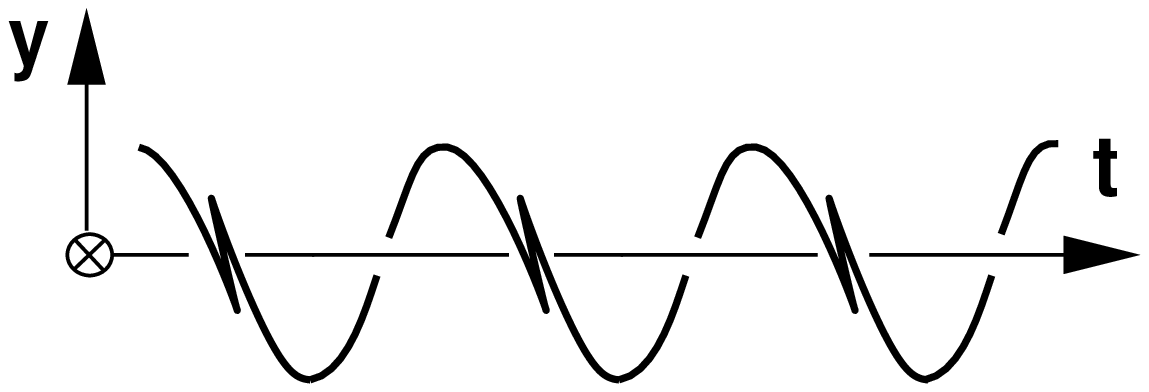}
\end{center}
\fignum Function $a_{0}(\s)$ and supporting curve for Example 2.
\end{figure}
\begin{figure}\label{f3}
\begin{center}
~\epsfysize=5cm\epsfxsize=5cm\epsffile{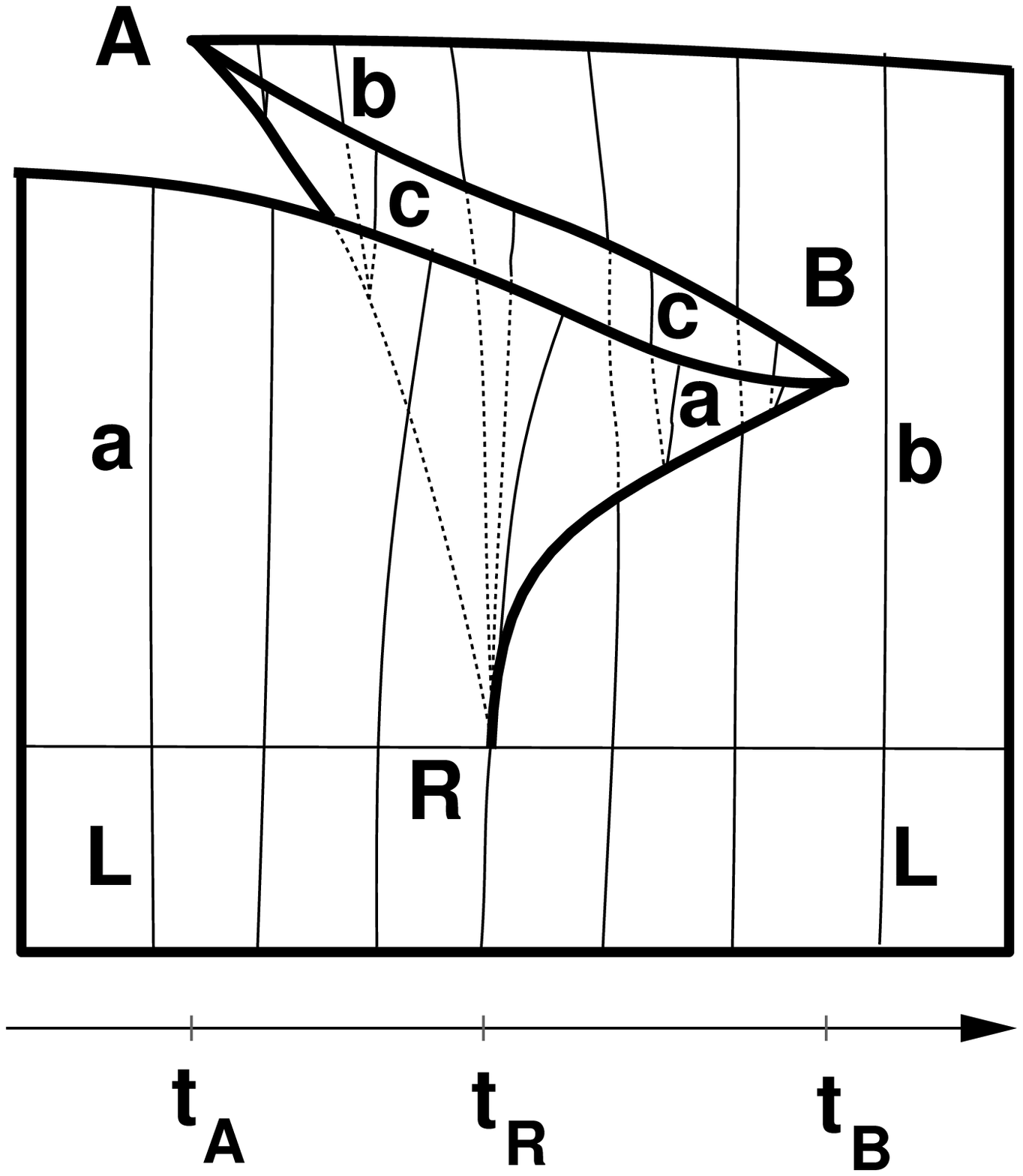}
\end{center}

\fignum Structure of the world sheet
in the vicinity of non-monotonous interval of the supporting curve.

\end{figure}}
\quad\quad
\parbox{6cm}{\begin{figure}\label{f4}
\begin{center}
~\epsfysize=6cm\epsfxsize=6cm\epsffile{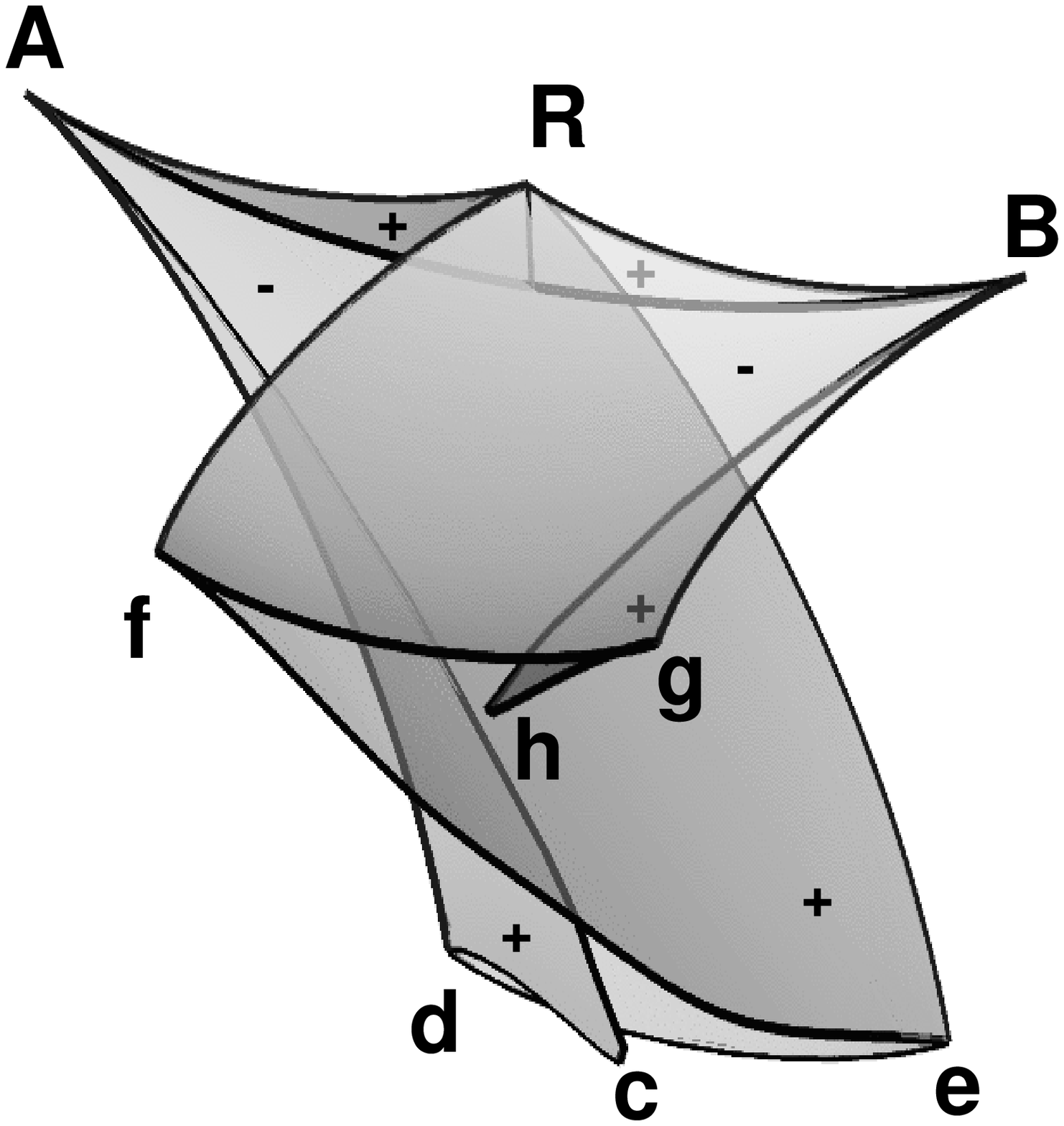}
\end{center}

\fignum Computer generated image of the world sheet.
This surface consists of 2 congruent patches $edAR$
and $fgBR$, separated by a partition 
$feR$. Supporting curve: $cABh$, cusp lines:
$fRAd$ and $gBRe$. $A,B$ -- creation and annihilation points,
$R=(A+B)/2$ -- recombination point.

\end{figure}}
\end{center}

% old picture: exotic_p6.tex
\vspace{2mm}

\begin{figure}\label{f7}
\parbox[c]{7cm}{
~\epsfysize=4cm\epsfxsize=7cm\epsffile{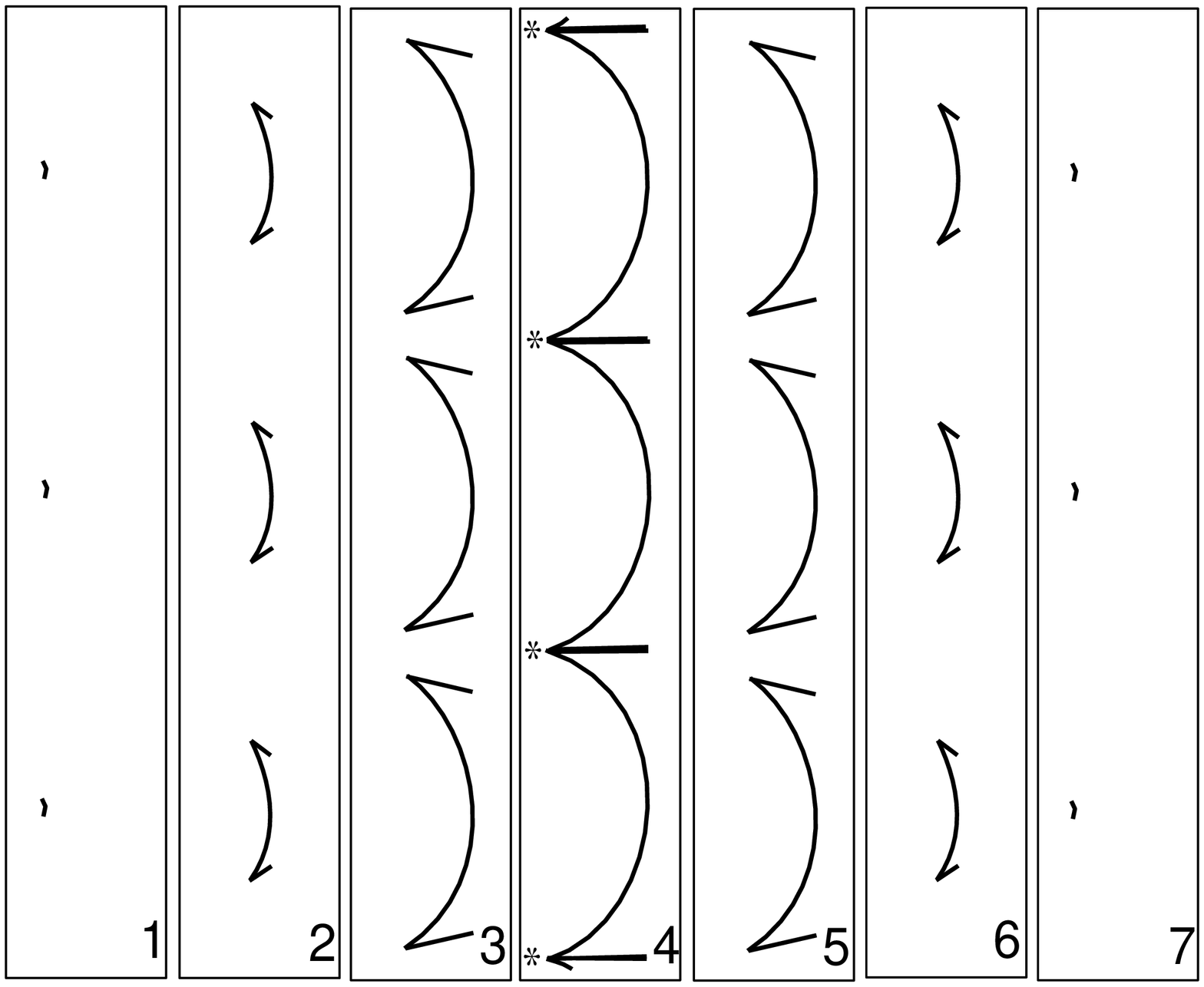}
}\quad\quad
\parbox[c]{6cm}{
\fignum
String dynamics for Example 4, observed in simultaneous reference frame.
An infinite number of strings, periodically located in one space
direction, simultaneously appear on the first frame. On the frame~4 
the strings recombine. On the last frame they simultaneously disappear. 
Further observation will show the empty space. 
}
\end{figure}

\begin{figure}\label{f8}
\parbox[c]{7cm}{
~\epsfysize=4cm\epsfxsize=6.5cm\epsffile{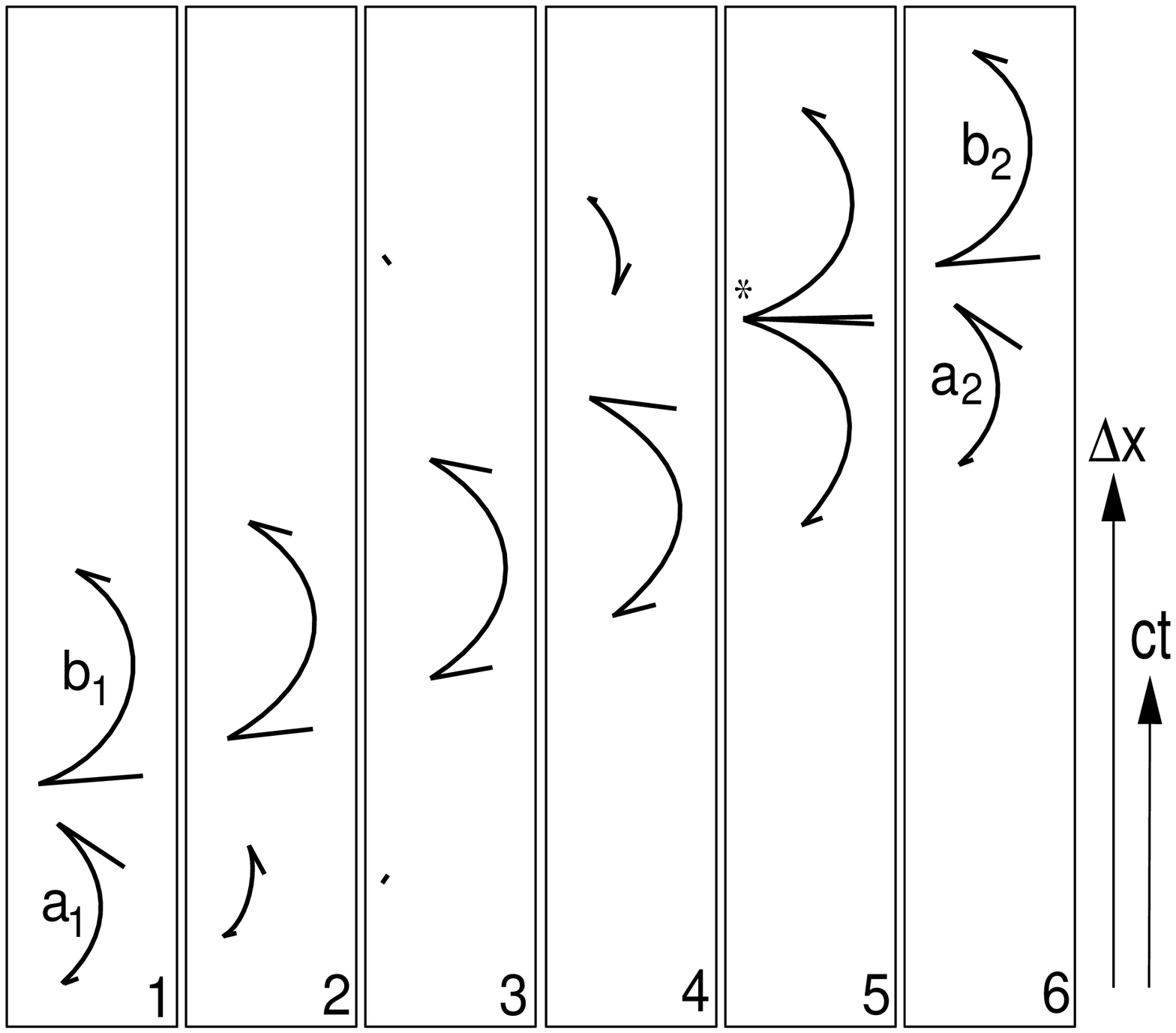}
}\quad\quad
\parbox[c]{6cm}{
\fignum
The same evolution, observed in other reference frame. Here we see the moving
string formation, which periodically repeats its form. On the first frame
strings $a_{1}$ and $b_{1}$ are presented. On the third frame
string $a_{1}$ disappears and string $b_{2}$ appears. 
The string $b_{1}$ deforms to the string $a_{2}$. 
The recombination occurs on the frame~5. 
The last frame is exact copy of the first one,
shifted by the distance $\Delta x$. This distance is greater than $ct$,
the path of light during the evolution.
}
\end{figure}

\vspace{3mm}\noindent\parbox[c]{9cm}{
\underline{{\it Example 4}}. 
$a_{0}(\s)=\cos\s$, see \fref{f7a}. For this supporting curve 
the period is space-like: $P_{\mu}=(0,\pi/2,0),\ P^{2}<0$. 
Correspondent world sheet is stretched not in time direction, but in 
the space one. There is no center-of-mass frame for this system. 
In such cases
one can use {\it simultaneous} reference frame, where the period has 
purely space direction (like in given example). 
In this frame the world sheet has finite
size in temporal direction, the evolution is restricted in time, see \fref{f7}. 
In other reference frames the world sheet is sloped to time axis,
the formation of strings, appearing in equal-time slice, 
has finite spatial sizes. It moves with the mean velocity,
equal to the slope $|\vec P|/P_{0}$, which is greater than 1, the light
velocity in our units. However, we will show in the next Section,
that all parts of this system do not exceed the light velocity. 
This motion is shown on \fref{f8}. 
It becomes clear from this figure, that there is no {\it direct} 
hyper-light transfer of the system, 
this transfer occurs through a sequence of creation, 
annihilation and recombination processes. Not initial string
but its exact copy approaches the finite point.
}\quad\quad\parbox[c]{4cm}{
\begin{figure}\label{f7a}
\begin{center}
~\epsfysize=6cm\epsfxsize=3cm\epsffile{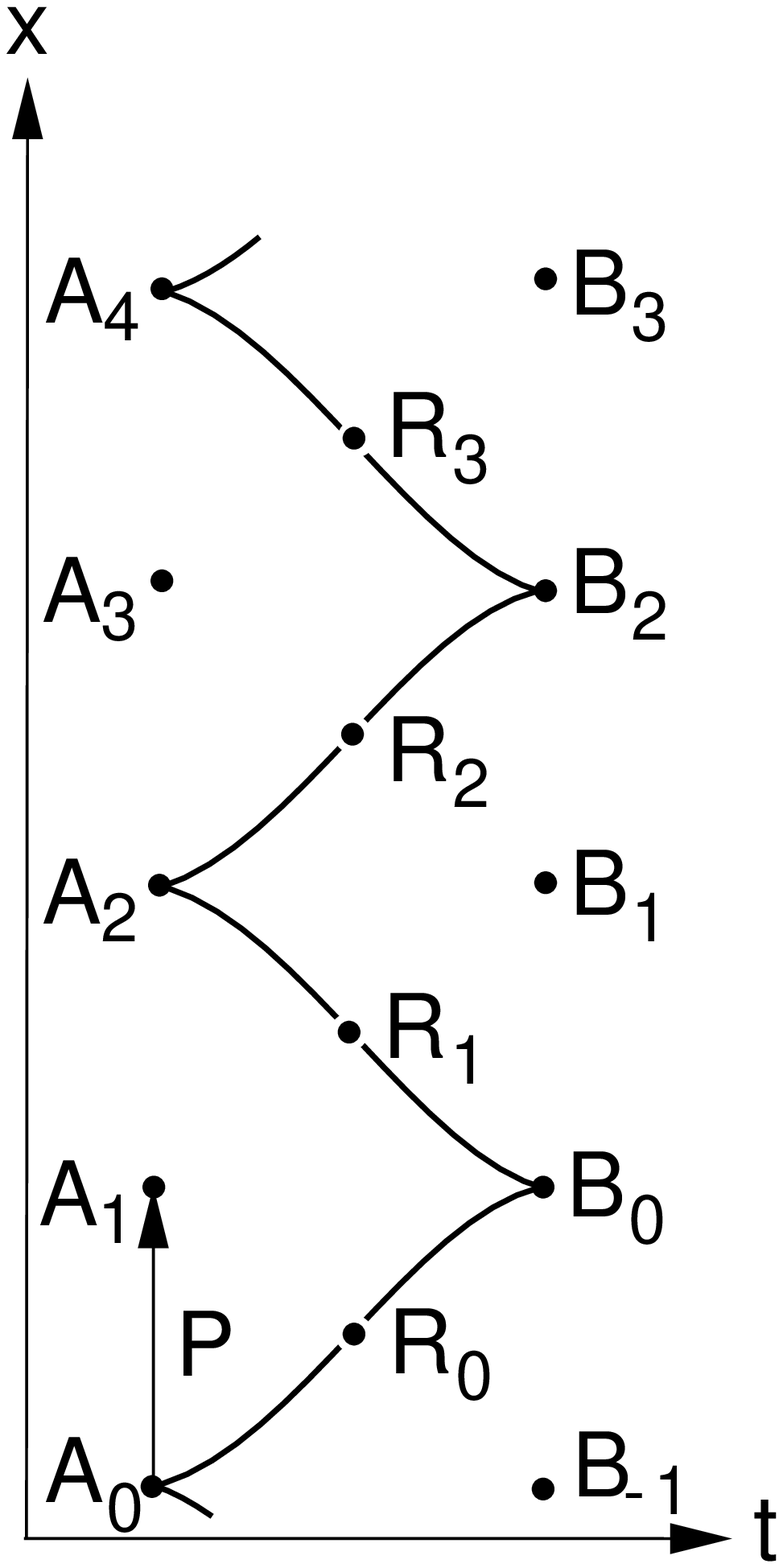}
\end{center}

\fignum Supporting curve for Example 4.
$A_{n}=A_{0}+Pn$ -- creation points, $B_{n}=B_{0}+Pn$ -- annihilation points, 
$R_{n}=(A_{0}+B_{0})/2+Pn$ -- recombination points.

\end{figure}
}

\noindent

\vspace{3mm}
\noindent

\paragraph*{1.2. Energetic flows on the world sheet.}\

Let's consider a connected component in equal-time slice
of the world sheet. Total momentum of this component
is given by the formula, derived in Appendix 2:
\begin{eqnarray}
&&\Delta P_{\mu}=\half(-Q_{\mu}(\s_{1})+Q_{\mu}(\s_{2}))
\vv_{(\s_{1},\s_{2})^{i}}^{(\s_{1},\s_{2})^{f}},\label{DPm}
\end{eqnarray}
where $(\s_{1},\s_{2})^{i,f}$ are respectively initial and final points
of the component on the parameters plane. For appearing/disappearing
component (e.g. short component, shown in bold on \fref{f5b})
we immediately have 
$$\Delta P_{\mu}=\half(-Q_{\mu}(\s_{1})+Q_{\mu}(\s_{2}))
\vv_{(\s,\s)}^{(\s',\s')}=0$$
(both ends of the string lie on the same edge of the world sheet).

For the permanent component, which starts and ends on different edges,
we have
$$\half(-Q_{\mu}(\s_{1})+Q_{\mu}(\s_{2}))
\vv_{(\s,\s)}^{(\s',\s'+2\pi)}=\half(-Q_{\mu}(\s')+Q_{\mu}(\s'+2\pi))=
P_{\mu},$$
the result is total momentum of the string.

Angular moment of the component is given by the formula
\begin{eqnarray}
&&\Delta M_{\mu\nu}=\quart Q_{1}^{[\mu}Q_{2}^{\nu]}
\vv_{(\s_{1},\s_{2})^{i}}^{(\s_{1},\s_{2})^{f}}+
\quart\left(-\int_{\s_{1}^{i}}^{\s_{1}^{f}}+\int_{\s_{2}^{i}}^{\s_{2}^{f}}
\right)Q^{[\mu}dQ^{\nu]}.\label{DM}
\end{eqnarray}
Here $Q^{\mu}_{i}=Q^{\mu}(\s_{i})$ 
and square brackets denote antisymmetrization
of indices: $A^{[\mu}B^{\nu]}=A^{\mu}B^{\nu}-A^{\nu}B^{\mu}$.
$\Delta M_{\mu\nu}$ is defined by two intervals of supporting curve: 
$\s\in[\s_{1}^{i},\s_{1}^{f}]\;\bigcup\;[\s_{2}^{i},\s_{2}^{f}]$.
It is not changed in reparametrizations of the supporting curve,
conserving boundary points of these intervals.
Generally, it depends on position of boundary points.

For appearing/disappearing components we have
$$\Delta M_{\mu\nu}=\quart Q_{1}^{[\mu}Q_{2}^{\nu]}
\vv_{(\s,\s)}^{(\s',\s')}+
\quart\left(-\int_{\s}^{\s'}+\int_{\s}^{\s'}
\right)Q^{[\mu}dQ^{\nu]}=0,$$
and for permanent component:
\begin{eqnarray}
&&M_{\mu\nu}=\quart Q_{1}^{[\mu}Q_{2}^{\nu]}
\vv_{(\s,\s)}^{(\s',\s'+2\pi)}+
\quart\left(-\int_{\s}^{\s'}+\int_{\s}^{\s'+2\pi}
\right)Q^{[\mu}dQ^{\nu]}=\nn\\
&&=\half Q^{[\mu}(\s')P^{\nu]}+\quart\int_{\s'}^{\s'+2\pi}
Q^{[\mu}dQ^{\nu]},\quad
{{dM_{\mu\nu}}\over{d\s'}}=\half(Q'^{[\mu}(\s')P^{\nu]}+
P^{[\mu}Q'^{\nu]}(\s'))=0.\nn
\end{eqnarray}
Thus, total momentum and angular moment for appearing/disappearing
components vanish, and creation/annihilation
processes do not violate the conservation laws.
For permanent component total momentum and angular moment
are presented as complete parametric invariants of the supporting curve,
independent on the position of ends ($\s,\s'$),
thus, they are conserved in evolution.

Vanishing of total energy for non-permanent component implies,
that density of energy is not everywhere positive on the string.
In Appendix 2 we derive a formula for linear density of energy
$dP_{0}/dl,\ dl=|d\vec x|$ -- element of length on equal-time slice:
\begin{eqnarray}
&&{{dP_{0}}\over{dl}}=\sgn(Q_{10}'Q_{20}')\cdot
\sqrt{{{2Q_{10}'Q_{20}'}\over{(Q_{1}'Q_{2}')}}}\ .\label{dp0dl}
\end{eqnarray}

\baselineskip=0.4\normalbaselineskip\footnotesize
\begin{center}
\parbox{0.95\textwidth}{
\underline{{\it Remark}}: argument of this square root is not negative,
because it can be presented in the form $2(n_{1}n_{2})^{-1}$,
where $n_{i\mu}=Q_{i\mu}'/Q_{i0}'$ are light-like vectors, 
directed into the future.}
\end{center}
\baselineskip=\normalbaselineskip\normalsize

For monotonous supporting curves ($Q_{0}'>0$) $dP_{0}/dl$
is always positive. For non-monotonous supporting curves
the density of energy on the string  changes its sign in passage through
the cusps (where the sign of $(Q_{10}'Q_{20}')$ is changed).
Parts of the world sheet, shown on \fref{f5b} in white and
double grey, have positive density of energy;
parts, shown in grey, correspond to negative density.
Signs of density of energy are also shown on the world sheet
\fref{f4}. In recombination process, shown on \fref{f5}, the density of energy
is positive on the segments $L,b,c,e$, and negative on the segments $a,d$.

\paragraph*{1.3. Gauges.}\ Virasoro constraints generate reparametrizations 
of supporting curve (see Appendix~1). Gauges to Virasoro constraints
select particular parametrization on this curve. For example,
the curve can be parameterized by cone time: $\s\sim(kQ)$ with
$k^{2}=0$, e.g. $k_{\mu}=(1,1,0,0...)$. In this parametrization
({\it light cone gauge}) Virasoro constraints can be explicitly
resolved:
$$Q_{\pm}=(Q_{0}\pm Q_{1})/\sqrt{2},\quad \vec Q_{\perp}=(Q_{2},...,Q_{d-1}),
\quad Q_{-}'=P_{-}/\pi>0,\quad 
Q_{+}'={\textstyle{{\pi}\over{2P_{-}}}}\;\vec Q_{\perp}'^{2}\geq0.$$

Another possibility is to parameterize the curve by the time in CMF
(for $P^{2}>0$): $\s\sim(PQ)$. Such gauge was originally considered by
Rohrlich~\cite{Rohrlich}. In this case $Q'^{0}=|\vec Q'|=\sqrt{P^{2}}/\pi>0$
($Q^{0}$ is a component of $Q_{\mu}$ along $P_{\mu}$, $\vec Q$ is a component,
orthogonal to $P_{\mu}$).

The both parametrizations can be introduced only on monotonous supporting 
curves, because we require that the component of $Q_{\mu}$ along 
gauge axis ($k_{\mu}$ or $P_{\mu}$) should be monotonous in $\s$.
As a result, exotic solutions are absent in light cone and Rohrlich's
gauges. Non-monotonous supporting curves can be considered in 
parametrization of another type, see Appendix~5.

\paragraph*{1.4. Regge condition.}\
For normal solutions the spatial components of angular moment
$M_{ij}$ in CMF satisfy the following inequality: $|M_{ij}|\leq P^{2}/2\pi$.
It can be proven by purely geometrical methods, see Appendix~3.
Alternatively, we can write the constraint
$L_{0}=0\ \Leftrightarrow\ P^{2}/2\pi=-\sum_{n>0}a_{\mu}^{n*}a_{\mu}^{n}$
in terms of variables $\alpha^{n}=(a_{i}^{n}+ia_{j}^{n})/\sqrt{2}$:
\begin{eqnarray}
&&{{P^{2}}\over{2\pi}}=\sum\limits_{n\neq0}\alpha^{n*}\alpha^{n}
+\sum\limits_{\twin{n>0}{l\neq i,j}}a_{l}^{n*}a_{l}^{n}-
\sum\limits_{n>0}a_{0}^{n*}a_{0}^{n},\quad
M_{ij}=\sum\limits_{n\neq0}{\textstyle{{1}\over{n}}}\;\alpha^{n*}\alpha^{n}.
\label{Regge_osc}
\end{eqnarray}
For normal solutions we can fix Rohrlich's gauge: $a_{0}^{n}=0$
at $n\neq0$, and for parametrically invariant variables $(M_{ij},P^{2}/2\pi)$
we will have: $|M_{ij}|=|\sum {\textstyle{{1}\over{n}}}\;\alpha^{n*}\alpha^{n}|
\leq\sum \alpha^{n*}\alpha^{n}\leq\sum \alpha^{n*}\alpha^{n}+\sum
a_{i}^{n*}a_{i}^{n}=P^{2}/2\pi$.

For exotic solutions the Rohrlich's gauge cannot be fixed,
and {\it negative} contribution of $a_{0}^{n}$-oscillators in
(\ref{Regge_osc}) can violate Regge-condition (and the condition $P^{2}\geq0$).
An explicit example:

\vspace{3mm}\noindent\underline{{\it Example 5:}}
$a_{\mu}(\s)=\alpha\cos2\s\;(1,\cos\s,\sin\s)$. In this case $P_{\mu}=0$
(supporting curve {\it is closed in Minkowsky space}, correspondent world 
sheet has finite sizes both in space and time directions).
$M_{12}=\half\oint Q_{1}dQ_{2}=-\pi\alpha^{2}/12\neq0$ (area, restricted by the curve
on $(12)$-plane, does not vanish).

\paragraph*{1.5. DDF variables.}\
In quantum theory the operators,
introduced by Del Giudice, Di Veccia and Fubini \cite{DDF}, 
play an important role. Here we will consider their classical analogs:

\vspace{-3mm}
\begin{eqnarray}
&&\vec A_{n}=\int_{0}^{2\pi}d\s\;\vec Q'_{\perp}
\exp in\;{{\pi Q_{-}}\over{P_{-}}}.\label{DDF_cl}
\end{eqnarray}

For normal solutions we can introduce the light cone gauge 
$\pi Q_{-}/P_{-}=\s$, in this case $\vec A_{n}$ coincide with 
transverse oscillator variables $\vec A_{n}=\int_{0}^{2\pi}d\s\;\vec Q'_{\perp}
e^{in\s}$ (formula
(\ref{DDF_cl}) gives parametrically-invariant expression for them).
For normal solutions the set of variables $\{\vec A_{n}\}$
defines the supporting curve uniquely (up to translations and 
reparametrizations).

For exotic solutions we cannot introduce the light cone gauge,
but variables $\vec A_{n}$ {\it are still defined}, both for time-like
and space-like $P_{\mu}$. In this case the supporting curve is not uniquely
defined by $\{\vec A_{n}\}$,
see Appendix~4. A deformation of supporting curve is possible, preserving
$\{\vec A_{n}\}$. 

\begin{center}
\begin{figure}\label{f9a}
\parbox{5cm}{
~\epsfysize=3.5cm\epsfxsize=4cm\epsffile{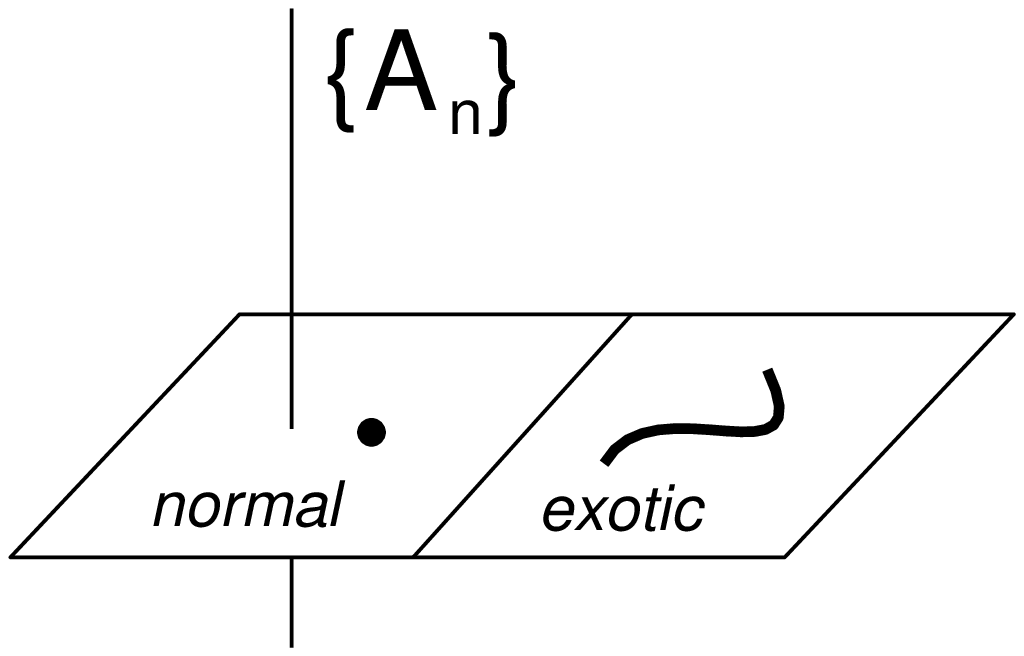}}
\quad\parbox{8cm}{
\fignum The set of DDF variables defines unique supporting curve
in sector of normal solutions, and a family of supporting curves
in exotic sector.
}
\end{figure}
\end{center}

%\baselineskip=0.4\normalbaselineskip\footnotesize
%\begin{center}
%\parbox{0.95\textwidth}{
\noindent\underline{{\it Remark.}} 

In theories with gauge symmetries the actual phase space is formed
by a complete set of gauge-invariant variables. Other variables 
can be ignored, because their variation transforms the
system along gauge equivalent states, indistinguishable physically.
All states of the system, correspondent to the same values of gauge
invariants, can be identified\footnote{For details of this procedure,
known as {\it gaugeless reduction}, see \cite{Blokhincev}.}.

In string theory variables $\{\vec A_{n}\}$ together with total momentum
and mean coordinate can be taken as such complete set {\it only for
normal solutions}. In this case other dynamical variables
influence only parametrization of supporting curve and do not influence
physical observables. 

For exotic solutions it is not so. A family of curves $Q_{\mu}(\s)$
exist, different by their shape, which correspond to the same set
$\{\vec A_{n}\}$, see \fref{f9a}. Identification of all states
with fixed $\{\vec A_{n}\}$ will lead to mixing of exotic and
normal states, i.e. identification of physically distinct states.
Later we will show, that such identification occurs in standard covariant
quantization of string theory.
%}
%\end{center}
%\baselineskip=\normalbaselineskip\normalsize

\section{Exotic solutions in Lagrangian theory}
Action of the string is defined as
\begin{eqnarray}
&&A=\int \L\; d\sigma_{1}d\sigma_{2},\quad
\L=\sqrt{(\df_{1}x\;\df_{2}x)^{2}-(\df_{1}x)^{2}(\df_{2}x)^{2}},\quad
\partial_{\alpha}=\partial/\partial\s_{\alpha}.\nn
\end{eqnarray}
The world sheets, considered in the previous Section,
have the following properties:

\vsp\noindent
1. They belong to the class ``open string'':
infinite band, mapped to Minkowsky space. 
The only difference from the usually considered world sheets is that
now this mapping is not trivial (has a fold, see figs.\ref{f3},\ref{f4}).

\vsp\noindent
2. The argument of the square root in $\L$ for these surfaces
is not negative: $(Q_{1}'Q_{2}')^{2}/16\geq0$. Such surfaces
are usually called time-like: in each regular point of the surface
one of two tangent vectors $Q_{1}'\pm Q_{2}'$ is time-like,
another one is space-like. This also implies, that velocity of points on 
the string does not exceed the light velocity\fproof{Considering equal-time
slices of the world sheet in the vicinity of regular point,
the action can be rewritten in the form $\int\sqrt{1-v_{\perp}^{2}}\;dtdl$,
where $dl$ is element of length on equal-time slice and $\vec v_{\perp}$
is a component of velocity, orthogonal to a string: only this component
has physical (parametrically invariant) sense. For time-like surfaces
$|\vec v_{\perp}|\leq1$.}

\vsp\noindent
3. The argument of the square root in $\L$ vanishes on the cusp lines
(figs.\ref{f4},\ref{f5b}): $(Q_{1}'Q_{2}')^{2}/16=0$. There are two
possibilities for a choice of sign of the square root in passage through
the cusp line:

\vspace{2mm}
a) Non-negative value is chosen for the square root: $\L=|Q_{1}'Q_{2}'|/4$.
In this case we can show (see Appendix~2), that Lagrange-Euler equations 
on each of two patches of the world sheet, separated by the cusp line,
are satisfied, but the boundary conditions on the cusp line are not:
momentum, flowing from one patch, is not equal to the momentum, entering
into another patch. In this case the considered world sheet 
{\it is not extremal surface} for action $A$.

\vspace{2mm}
b) The square root changes sign in passage through the cusp line:
$\L=(Q_{1}'Q_{2}')/4$, $\sgn(Q_{1}'Q_{2}')=\sgn(Q_{10}'Q_{20}')$
(parts, marked by $+$ and $-$ on \fref{f4}, contribute to the action
with opposite signs). The string, whose Lagrangian changes sign
in passage through singular points (``polarized strings'')
were initially considered in work \cite{Zhelt}. 
In this case the Lagrange-Euler equations and boundary conditions are 
satisfied, and the extremum of action is reached on considered surface.

\vspace{2mm}
Thus, exotic solutions of covariant Hamiltonian theory correspond to
solutions of Lagrangian theory of type $b$. In the case $a$ such
solutions should be excluded.

%\baselineskip=0.4\normalbaselineskip\footnotesize
%\begin{center}
%\parbox{0.95\textwidth}{

\vspace{3mm}
\noindent\underline{{\it Remark.}} 
String theory is usually considered in conformal coordinates
$(\tau,\s)$, related with isotropic ones $(\s_{1},\s_{2})$
by expression $\s_{1,2}=\tau\pm\s$. In these coordinates
$\dot x=\df x/\df\tau=(Q_{1}'+Q_{2}')/2,\ x'=\df x/\df\s=(Q_{1}'-Q_{2}')/2,\ 
\dot xx'=\dot x^{2}+x'^{2}=0$. One can consider a family of straight 
lines $\tau=(\s_{1}+\s_{2})/2=Const$ on parameters plane \fref{f5b} 
(instead of equal-time slices)
and obtain correspondent foliation of the world sheet by strings
in Minkowsky space (in this case the strings are connected curves).
Namely this foliation occurs in action $A$, reformulated in terms
of coordinates $(\tau,\s)$. For normal solutions
(monotonous supporting curves) in each regular point
of the world sheet the vector $x'$ is space-like, $\dot x$ is time-like
and directed to the future: $x'^{2}<0,\ \dot x^{2}>0,\ \dot x_{0}>0$.
For exotic solutions it is not so: in white regions of \fref{f5b}
$x'^{2}<0,\ \dot x^{2}>0,\ \dot x_{0}>0$; in grey regions
$x'^{2}>0,\ \dot x^{2}<0$; in double grey regions 
$x'^{2}<0,\ \dot x^{2}>0,$ but $\dot x_{0}<0$. 
Usually such solutions are rejected from consideration on early stage
of theory construction, by explicit requirement 
$x'^{2}<0,\ \dot x^{2}>0,\ \dot x_{0}>0$, see e.g. \cite{Brink}.
This requirement, however, remains {\it unused} 
in further development of the theory. To exclude the exotic solutions
from Hamiltonian theory, one should explicitly impose the requirement
$a_{0}(\s)>0\ \forall\s$. In the space of oscillator variables
$\{a_{0}^{n}\}$ this requirement defines a region (linearly connected and 
compact for each fixed $a_{0}^{0}=P_{0}/\sqrt{\pi}$). Boundary of this region
is composed of patches, whose explicit expression can be found
in exclusion of variable $z=e^{i\s}$ from the system
$a_{0}(z)=0,\ a_{0}'(z)=0$. For finite number of excited oscillators
this expression can be written as polynomial of oscillator variables
\cite{Groebner}, with the degree, increasing together with the number of
excited oscillators. Needless to say, that determination of this region
is a hard problem. Until it will not be solved,
exotic solutions exist in Hamiltonian theory, 
and should also be present on a quantum level.
%}
%\end{center}
%\baselineskip=\normalbaselineskip\normalsize

\section{Exotic solutions in quantum theory}

Standard covariant quantization of string theory works in 26-dimensional
Minkowsky space. Physical space of states is defined as
$L^{n}|phys\ran=0,\ n\geq0$ (only a half of constraints is imposed).
In this theory the property $a_{0}(\s)>0$ cannot be directly tested,
because operator $a_{0}(\s)$ does not commute with $L^{n}$
and, consequently, does not take definite values on the physical vectors.
Only indirect methods can be used for study of exotic sector in
quantum theory.

Square of string's mass is defined by the expression
$P^{2}/2\pi=\sum a^{n+}_{i}a^{n}_{i}-\sum a^{n+}_{0}a^{n}_{0}$.
In classical string theory the excitation of $a^{n}_{0}$-oscillators
gives negative contribution to $P^{2}/2\pi$ and can lead to
arbitrarily great negative square of mass.

In standard covariant quantum string theory the vacuum is 
annihilated by operators $a_{\mu}^{n}|0\ran=0,\ n>0$.
Time component of oscillator variables has a commutator
$[a_{0}^{n},a_{0}^{n+}]=-n$, opposite to the commutator of space
components: $[a_{i}^{n},a_{i}^{n+}]=n$.
Operators $a_{0}^{n+}$ create from vacuum the states 
$(a_{0}^{n+})^{k}|0\ran$, for which correspondent
occupation number operator $a^{n+}_{0}a^{n}_{0}$ 
takes {\it negative} eigenvalues,
e.g. $(a_{0}^{1+}a_{0}^{1})a_{0}^{1+}|0\ran=-a_{0}^{1+}|0\ran$
(for odd $k$ the states $(a_{0}^{n+})^{k}|0\ran$ have also negative norm).
As a result, the excitation of $a^{n}_{0}$-oscillators
in quantum theory gives positive contribution to $P^{2}/2\pi$,
and square of mass is not negative\footnote{Exception is
a single tachionic state, which appears after the redefinition 
$P^{2}/2\pi\to P^{2}/2\pi-1$ (introduction of intercept).}.
Analogously, considering the expression (\ref{Regge_osc}),
we can prove that Regge condition in quantum theory is not violated.

Such exclusion of tachionic and non-Regge regions,
present in exotic sector on classical level, has absolutely different
mechanism, than the requirement $a_{0}(\s)>0$. 
Actually, this is not
the exclusion of these states, but {\it another definition} 
of their square of mass. This redefinition
maps the regions in the phase space,
violating positiveness of $P^{2}$ and Regge condition,
to the regions of spectrum, where these conditions 
are not violated.

Now let´s consider DDF states $\prod_{n<0,i}(A_{ni})^{k_{ni}}|0\ran$,
operators $A_{ni}$ are defined by expression (\ref{DDF_cl}) 
(no special ordering
is required). On these states $a_{-}^{n}|DDF\ran=0,\ n>0$, 
classically this corresponds to
$a_{-}(\s)=Const$, i.e. light cone parametrization.
Due to the theorem \cite{DDF_theorem}, any physical state can be 
presented in a form
$|phys\ran=|DDF\ran+|sp\ran$, where $|DDF\ran$ belongs to the space,
covered by DDF-states, and $|sp\ran$ is a spurious state, which has 
zero norm and is orthogonal to all states in the physical space.
First of all, note that exotic solutions necessarily have non-constant
$a_{-}(\s)$. Consequently, they cannot correspond to pure DDF-states,
but should have non-zero spurious component. Then, actual space of states
is obtained in factorization of the physical space by spurious states:
$|DDF\ran+|sp\ran\sim|DDF\ran$. This factorization is equivalent to one
described in subsection 1.5, which leads to identification of normal and exotic
solutions.

%\baselineskip=0.4\normalbaselineskip\footnotesize
%\begin{center}
%\parbox{0.95\textwidth}{
\vspace{3mm}
\noindent\underline{{\it Remarks.}} 

\vsp\noindent
1. Such situation occurs in standard representation of quantum string theory.
In work \cite{indef} alternative representation was considered. It was shown,
that in pseudo-Euclidean space, containing equal number of spatial
and temporal directions (e.g. in $d=3+3$), quantum string theory
can be represented in positively defined extended space of states, 
where correspondent Virasoro algebra has {\it zero central charge}. In this theory
the square of string's mass has the expression, analogous to (\ref{Regge_osc}),
but all occupation number operators in this theory take positive eigenvalues.
As a result, tachionic regions of classical theory are preserved
in quantization.

\vsp\noindent
2. Light cone and Rohrlich's gauges exclude exotic solutions on the classical level.
Correspondent quantum theory also does not contain the exotic solutions.
It is well known for standard light cone quantization, applicable in 26 dimensions.
In works [10-12] %\cite{straight,par6,ax} 
special subsets in the phase space
of 4-dimensional open string theory were considered, for which the quantization
in Rohrlich's gauge and in (slightly modified) light cone
gauge has no anomalies. The obtained spectra also do not contain the exotic sector.

Quantum string theory in the gauge, described in Appendix~5,
is anomaly-free as well. It contains exotic solutions classically and 
preserves them on the quantum level.
%}
%\end{center}
%\baselineskip=\normalbaselineskip\normalsize

\section*{Conclusion}

We considered a class of solutions in Nambu-Goto theory of open string,
which correspond to the world sheets, mapped in Minkowsky space with a fold.

\vsp\noindent
1. Such solutions exist in covariant Hamiltonian formulation of the theory.
Typical processes for these solutions are creation of strings from vacuum,
their recombination and annihilation. 
These solutions can violate positiveness of square of mass and Regge condition
$|M_{ij}|\leq P^{2}/2\pi$. Such solutions are present, until a special
requirement $a_{0}(\s)>0$ will be explicitly imposed.

In non-covariant formulations the exotic solutions are absent for commonly used
gauges (Rohrlich's and light cone) and present for a gauge of special type, 
described here in Appendix~5.

\vsp\noindent
2. In Lagrangian formulation a behavior of the square root in Lagrangian 
near its branching point plays defining role. If positive sign of the 
square root
is chosen, the exotic solutions are absent (and should be excluded from
correspondent Hamiltonian theory). If Lagrangian changes its sign in passage
through the branching point, the exotic solutions are present.

\vsp\noindent
3. In covariant quantum theory imposition of the requirement $a_{0}(\s)>0$
is a subtle task. Until this requirement is not imposed, the exotic solutions 
are present.

Standard covariant quantization of string theory operates in the space 
of states
with indefinite metric. This leads to redefinition of string's square of mass,
which maps the regions of classical theory with $P^{2}<0$ and 
$|M_{ij}|>P^{2}/2\pi$
to the regions of quantum spectra with $P^{2}>0$ and $|M_{ij}|<P^{2}/2\pi$.
After that redefinition, the exotic solutions correspond to physical states
$|DDF\ran+|sp\ran$ with non-zero spurious component. Factorization 
of the physical
space by spurious states leads to mixing of exotic and normal solutions.

An example of covariant quantization of string theory in positively defined
extended space of states is known, which contains exotic solutions in spectrum.

In non-covariant quantization the gauges, excluding exotic solutions 
classically,
do not have them on the quantum level. The gauge, considered in Appendix~5,
contains the exotic solutions classically and preserves them on the quantum 
level.

\vsp
Finally, we want to describe the following problem, which we foresee in
further development of string theory. One day somebody
can find quantum representation of string theory in $d=3+1$,
which will be free of anomalies and will operate in positively defined
extended space of states. This theory, however, necessarily should contain
exotic states. Their exclusion is possible only in specific gauges,
which do not contain exotic solutions classically. This requirement
narrows the class of admissible gauges.

\paragraph*{Acknowledgment.}
We are indebted to Prof. George P. Pronko and Dr. Martin G\"obel
for the support and encouragement of our work on this subject.

The images, presented in this work, are constructed by computer program
\cite{vis}, which can create static 3D models of the world sheets
and present string dynamics as a film. This program is developed
in the frames of a project ``Visualization of complex physical phenomena
and mathematical objects in virtual environment'', supported by INTAS 96-0778 
grant. 

\baselineskip=0.4\normalbaselineskip\footnotesize

\section*{Appendix 1. Geometrical reconstruction of the world sheets}

\noindent
Properties 1,2 of supporting curve follow from its definition.
Property 3 follows from 4 and 2. Let's prove the property~4:
\begin{eqnarray}
&&x_{\mu}(\s,\tau)=(Q_{\mu}(\s_{1})+Q_{\mu}(\s_{2}))/2,\quad 
\s_{1,2}=\tau\pm\s.\label{sheet}
\end{eqnarray}
This formula was obtained in \cite{zone} by direct solution of Hamiltonian 
equations in $(x,p)$-representation. Here we will reproduce the proof 
of this formula in oscillator representation.

Coordinates and momenta of the string are defined by expressions\footnote{
See e.g.\cite{Brink}. Difference of notations: $a_{\mu}^{n}$ 
in our work corresponds to $i\sqrt{n}a_{\mu}^{n*}$ in \cite{Brink} ($n>0$).}
\begin{eqnarray}
&&x_{\mu}(\sigma)=X_{\mu}+{\textstyle{{1}\over{\sqrt{\pi}}}}
\sum\limits_{n\neq0}{\textstyle{{a_{\mu}^{n}}\over{in}}}\cos n\sigma,\quad
p_{\mu}(\sigma)={\textstyle{{1}\over{\sqrt{\pi}}}}
\sum\limits_{n}a_{\mu}^{n}\cos n\sigma,\nn
\end{eqnarray}
so that formulae (\ref{Qini0})-(\ref{Qosc}) are valid.
Poisson brackets for canonical variables:
\begin{eqnarray}
&&\{a_{\mu}^{n},a_{\nu}^{k}\}=in\;g_{\mu\nu}\;\delta^{k,-n}, 
\quad \{X_{\mu},P_{\nu}\}=g_{\mu\nu}.\nn
\end{eqnarray}
Hamiltonian of the system is an arbitrary linear combination of Virasoro
constraints: $H=\sum c^{k}L^{k}\ (c^{k*}=c^{-k})$.
Coefficients $c^{k}$ influence only parametrization of the world sheet.
The choice $H=L^{0}$ corresponds to conformal parametrization (where
$(x'\pm\dot x)^{2}=0$). This Hamiltonian generates phase rotations
$a_{\mu}^{n}(\tau)=a_{\mu}^{n}(0)e^{in\tau}$ and shifts 
$X_{\mu}(\tau)=X_{\mu}(0)+(P_{\mu}/\pi)\tau $.
Using (\ref{Qosc}), we see that the evolution of function $Q_{\mu}(\sigma)$ 
is the shift of its argument: $Q_{\mu}(\sigma,\tau)
=Q_{\mu}(\tau+\sigma,0)$. Then, using (\ref{Qini}), we have the 
following evolution for coordinates and momenta:
\begin{eqnarray}
&&x_{\mu}(\sigma,\tau)=(Q_{\mu}(\tau+\sigma,0)+Q_{\mu}(\tau-\sigma,0))/2,\quad
p_{\mu}(\sigma,\tau)=(Q'_{\mu}(\tau+\sigma,0)+Q'_{\mu}(\tau-\sigma,0))/2.\nn
\end{eqnarray}
Introducing isotropic coordinates $\sigma_{1,2}=\tau\pm\sigma$,
obtain formula (\ref{sheet}).

\vspace{3mm}
\noindent\underline{{\it Remark:}}
Poisson brackets for $Q_{\mu}(\s)$:
$\{Q_{\mu}(\s),Q_{\nu}(\tilde\s)\}=-2g_{\mu\nu}\vartheta(\s-\tilde\s),$
where $\vartheta(\s)=[\s /2\pi]+\half$, $[x]$ is integer part of $x$, 
a derivative $\vartheta(\s)'=\Delta(\s)$ is periodical delta-function.
As a result, we have $\{Q_{\mu}(\s),Q'^{2}(\tilde\s)/4\}=
\Delta(\s-\tilde\s)Q'_{\mu}(\s),$ and for 
$H=\int d\s\;F(\s)Q'^{2}(\s)/4:\quad 
\dot Q_{\mu}(\s)=\{Q_{\mu}(\s),H\}=F(\s)Q'_{\mu}(\s)$,
linear combinations of constraints generate shifts of points 
in tangent direction to the supporting curve, or equivalently --
reparametrizations of this curve. 

Reduced phase space of string, obtained in factorization of the phase space
by the action of gauge group, is actually a set of all possible supporting 
curves, which are considered as geometric images, without respect to their
parametrization (two different parametrizations of the curve correspond to
the same point of the reduced phase space). 
All physical observables in string theory are parametric invariants 
of supporting curve. The world sheet is also reconstructed by the supporting 
curve in parametrically invariant way, see \fref{f1}. 

\section*{Appendix 2. Densities of momentum and angular moment}

\noindent
Action of the string is equal to an area of the world sheet
\begin{eqnarray}
&&A=\int \L\; d\sigma_{1}d\sigma_{2},\quad
\L=\sqrt{(\df_{1}x\;\df_{2}x)^{2}-(\df_{1}x)^{2}(\df_{2}x)^{2}}
=\sqrt{-\half \L^{\mu\nu}\L^{\mu\nu}},\quad
\L^{\mu\nu}=\epsilon_{\alpha\beta}\;\partial_{\alpha}x^{\mu}
\partial_{\beta}x^{\nu},\nn
\end{eqnarray}
where $\epsilon_{12}=-\epsilon_{21}=1,\ \epsilon_{11}=\epsilon_{22}=0.$
$\L^{\mu\nu}d\sigma_{1}d\sigma_{2}$ is tensor element of area.

\begin{figure}\label{fA21}
\parbox{5cm}
{~\epsfysize=2cm\epsfxsize=3.8cm\epsffile{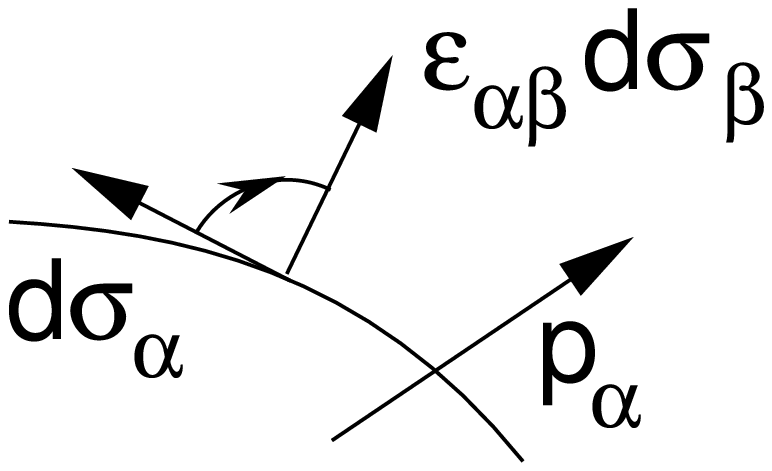}}\quad
\parbox{8cm}
{\fignum Tangent and normal elements to the curve on parameters plane.}
\end{figure}

Condition for extremum of action has a form:
\begin{equation}
\partial_{\alpha}p_{\alpha}^{\mu}=0,\quad 
p_{\alpha}^{\mu}={{\delta A}\over{\delta\;\partial_{\alpha}x^{\mu}}}
=-{{\L^{\mu\nu}}\over{\L}}\;\epsilon_{\alpha\beta}\;\partial_{\beta}x^{\nu},
\label{cons}
\end{equation}
$p_{\alpha}^{\mu}$ is momentum density, flowing through a section
$\sigma_{\alpha}=Const$. Integral form of this condition:
$\oint_{C}p_{\alpha}^{\mu}\;\epsilon_{\alpha\beta}\; d\sigma_{\beta}=0$
has a sense of momentum conservation law: total momentum, flowing through a
closed contour $C$ on the world sheet, vanishes 
(here $d\s_{\alpha}=(d\s_{1},d\s_{2})$ is tangent element to the curve $C$,
$\epsilon_{\alpha\beta}d\s_{\beta}=(d\s_{2},-d\s_{1})$ is the normal element,
see \fref{fA21}).

\vspace{5mm}

\vspace{2mm}

\noindent\underline{{\it Remarks.}}

\vsp\noindent
1. (Conventions). Usually the action of the string is 
multiplied by a factor $(2\pi\alpha')^{-1}$,
$\alpha'$ is a dimensional constant, defining the tension of the string.
In our work we choose the system of units $2\pi\alpha'=1$.

Also the sign {\it minus} is usually introduced in the action of the string,
by analogy with the mechanics of point particle, where
$A=-ms$, $s$ is interval of the world line.
This sign is actually needed only at the choice of metric
with signature $(-++\cdots)$. For this choice it provides
positiveness of energy (time component of {\it covariant}
vector $p_{\mu}=g_{\mu\nu}\df L/\df\dot x_{\nu}$),
see table below
$(\vec v=d\vec x/dx_{0},\ \gamma=(1-v^{2})^{-1/2})$:

\begin{center}
\begin{tabular}{cccc}\hline
signature&$L$&$p^{\mu}=\df L/\df\dot x_{\mu}$&$p_{\mu}=g_{\mu\nu}p^{\nu}$
\\\hline
$-++\cdots$&$-m\sqrt{-\dot x^{2}}$&$m\gamma(-1,\vec v)$&$m\gamma(1,\vec v)$
\\\hline
$+--\cdots$&$m\sqrt{\dot x^{2}}$&$m\gamma(1,-\vec v)$&$m\gamma(1,\vec v)$
\\\hline
\end{tabular}
\end{center}

For signature $(+--\cdots)$, adopted in our work,
positive Lagrangian corresponds to positive energy.
Analogously, the choice of sign {\it plus} in string's action
corresponds to positive density of energy for normal solutions
(see below).

\vsp\noindent
2. In the points of the world sheet, where $\L=0$, $p_{\alpha}^{\mu}$
has a singularity. Such cases are often rejected from consideration
(see e.g. \cite{Brink}). This rejection has no ground.
It is shown in \cite{sing}, that points with $\L=0$ for the theory
in 3- and 4-dimensional Minkowsky space are topologically stable.
They do not disappear in small deformations of the world sheet.
Such singular cases occupy volumes in the phase space (are not rare).
For the world sheets, considering in our work, $\L=0$ on cusp lines
(where $\df_{1}x$ or $\df_{2}x$ vanishes). Thus, such singular cases
should be considered in details.

\begin{figure}\label{fA22}
\parbox{4cm}
{~\epsfysize=2.5cm\epsfxsize=3.75cm\epsffile{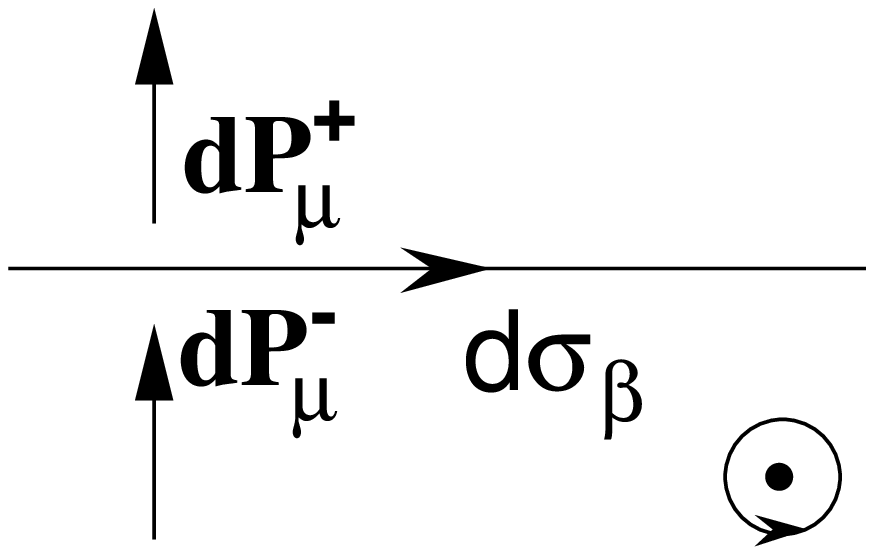}}\quad
\fignum Flows of momentum through the line $\L=0$.
\end{figure}

Let's subdivide the world sheet into patches with $\L\neq0$,
separated by lines with $\L=0$ (isolated points with $\L=0$
are also rejected from the patches). Variation of action
in inner regions of $\L\neq0$ patches gives equations (\ref{cons})
for these regions. Due to these equations, variation of action
is reduced to boundary terms:
$\delta A=\sum_{i}\int_{B_{i}}\delta x_{\mu}(dP_{\mu}^{+}-dP_{\mu}^{-})$,
where $dP_{\mu}=p_{\alpha}^{\mu}\epsilon_{\alpha\beta}d\sigma_{\beta}$
is a differential flow of momentum through the element $d\sigma_{\beta}$,
see \fref{fA22}. This expression is well defined, only if 
$dP_{\mu}$ has finite limits on the boundary, correspondingly 
$dP_{\mu}^{+}$ and $dP_{\mu}^{-}$ for each of two patches,
adjacent to $d\sigma_{\beta}$.

Annulation $\delta A=0$ for any $\delta x$ implies,
that $dP_{\mu}^{+}=dP_{\mu}^{-}$, limits of differential flows
should coincide on the boundary. 
In other words: momentum, flowing from one patch, should be equal
to momentum, entering into another patch, so that momentum conservation
is satisfied globally on the world sheet\footnote{
In this work we actually consider only such world sheets,
for which the nominator in $p_{\alpha}^{\mu}$ vanishes simultaneously 
with the denominator, and resolution of this $0/0$ ambiguity
gives smooth ($C^{\infty}$) function $p_{\alpha}^{\mu}(\s_{1},\s_{2})$.
For such world sheets the limits $dP_{\mu}^{\pm}$ automatically coincide,
and the equations $\partial_{\alpha}p_{\alpha}^{\mu}=0$
have sense everywhere on the world sheet.}. (For isolated singular
point the momentum, flowing through a closed contour around this point
should vanish. Analogously, boundary condition on a free edge of the 
world sheet has a form $dP_{\mu}=0$ -- flow of momentum through the free edge
should vanish.)

\vspace{5mm}

\vspace{5mm}

Let's substitute the representation of the world sheet
$x(\s_{1},\s_{2})=(Q(\s_{1})+Q(\s_{2}))/2$ into~(\ref{cons}):
\begin{eqnarray}
&&\df_{1,2}x=Q_{1,2}'/2,\ \ \L_{\mu\nu}=Q'_{1[\mu}Q'_{2\nu]}/4,\ \ 
-\half\L_{\mu\nu}\L_{\mu\nu}=(Q'_{1}Q'_{2})^{2}/16,\nn\\ 
&&\L=(Q'_{1}Q'_{2})/4,\ \ p_{1,2}^{\mu}=Q'_{\mu}(\s_{2,1})/2,\nn
\end{eqnarray}
consequently, the equations (\ref{cons}) are valid: 
$\df_{1}p_{1}=\df_{2}p_{2}=0$. Boundary conditions on a free edge
are also satisfied (e.g. for the edge $\s_{1}=\s_{2}$
we have $p_{1}=p_{2},\ d\s_{1}=d\s_{2}\ \Rightarrow\ 
dP=p_{1}d\s_{2}-p_{2}d\s_{1}=0$). 
Thus, the world sheets of the form
$x=(Q_{1}+Q_{2})/2$ satisfy Lagrange-Euler equations and boundary conditions,
also in the case, if the supporting curve $Q(\s)$ is not
monotonous. 

\vsp
\noindent\underline{{\it Remark:}}
this result is valid in the case of Lagrangian theory 
of type $b$, see Section~2. For the theory of type $a$ we obtain 
$\L=|Q'_{1}Q'_{2}|/4,\ p_{1,2}^{\mu}=Q'_{\mu}(\s_{2,1})/2\cdot
\sgn(Q_{10}'Q_{20}')$, function $p_{\alpha}^{\mu}$ is discontinuous on
cusp lines \fref{f5b}. In this case the flows $dP^{\pm}_{\mu}$ do not coincide 
on the cusp lines (e.g. for $\s_{1}=\s_{A,B}+2\pi n$ we have
$dP^{+}-dP^{-}=\pm Q'_{2}d\s_{2}\neq0$). 
Further consideration is concerned to the theory of type $b$.

The momentum, flowing through the element $(d\s_{1},d\s_{2})$,
is given by an expression
\begin{eqnarray}
&&dP^{\mu}=p_{\alpha}^{\mu}\epsilon_{\alpha\beta}d\sigma_{\beta}=
(-Q'_{1}d\s_{1}+Q'_{2}d\s_{2})^{\mu}/2.\label{dPm}
\end{eqnarray}
Calculating $\int dP^{\mu}$ along the curve, connecting points
$(\s_{1},\s_{2})^{i}$ and $(\s_{1},\s_{2})^{f}$, obtain
formula~(\ref{DPm}).

Density of angular moment:
\begin{eqnarray}
&&dM_{\mu\nu}=x_{[\mu}dP_{\nu]}=\quart(Q_{1}+Q_{2})_{[\mu}
d(-Q_{1}+Q_{2})_{\nu]}.\nn
\end{eqnarray}
Integrating this density along the curve, have
\begin{eqnarray}
&&\Delta M_{\mu\nu}=\quart Q_{1}^{[\mu}Q_{2}^{\nu]}
\vv_{(\s_{1},\s_{2})^{i}}^{(\s_{1},\s_{2})^{f}}+
\quart\int_{(\s_{1},\s_{2})^{i}}^{(\s_{1},\s_{2})^{f}}
(-Q_{1}^{[\mu}dQ_{1}^{\nu]}+Q_{2}^{[\mu}dQ_{2}^{\nu]}).\nn
\end{eqnarray}
This formula can be identically rewritten to the form (\ref{DM}).

Let's derive a formula (\ref{dp0dl}) for linear density of energy
on the string. Let's consider equal-time slice of the world sheet:
\begin{eqnarray}
&&dx_{0}(\s_{1},\s_{2})=(Q_{10}'d\s_{1}+Q_{20}'d\s_{2})/2=0\
\Rightarrow\ d\s_{1}=-{{Q_{20}'}\over{Q_{10}'}}d\s_{2}.\label{ds12}
\end{eqnarray}
From (\ref{dPm}): $dP_{0}=(-Q_{10}'d\s_{1}+Q_{20}'d\s_{2})/2=Q_{20}'d\s_{2}$.
Element of length on the string:
\begin{eqnarray}
&&d\vec x(\s_{1},\s_{2})=(\vec Q_{1}'d\s_{1}+\vec Q_{2}'d\s_{2})/2
=\left(-\vec Q_{1}'{{Q_{20}'}\over{Q_{10}'}}+\vec Q_{2}'\right)d\s_{2},\
dl=|d\vec x|=\half\left|-\vec Q_{1}'{{Q_{20}'}\over{Q_{10}'}}+\vec Q_{2}'
\right|\;|d\s_{2}|,\nn
\end{eqnarray}
or after elementary transformations: 
$dl=\sqrt{(Q'_{1}Q'_{2})Q_{20}'/2Q_{10}'}\cdot|d\s_{2}|$\quad
(argument of the square root is not negative).

The linear density of energy:
\begin{eqnarray}
&&{{dP_{0}}\over{dl}}=\sqrt{{{2Q_{10}'}\over{Q_{20}'(Q'_{1}Q'_{2})}}}\cdot
Q_{20}'{{d\s_{2}}\over{|d\s_{2}|}}.\label{dp0dl1}
\end{eqnarray}

To find the sign of $d\s_{2}$, let's consider normal element
$\epsilon_{\alpha\beta}d\sigma_{\beta}=(d\s_{2},-d\s_{1})$,
see \fref{fA21}. Expression (\ref{dPm}) actually represents
the flow of momentum in the direction of normal element.
In our case the normal element should correspond to the Lorentz
vector $(Q_{1}'d\s_{2}-Q_{2}'d\s_{1})$, 
{\it directed into the future}. In this case we calculate the flow 
of momentum through the equal-time slice from past to the future.
This flow is usually identified with total momentum of the string.

Thus, $Q_{10}'d\s_{2}-Q_{20}'d\s_{1}>0\ \Rightarrow\ 
(Q_{10}'+(Q_{20}')^{2}/Q_{10}')d\s_{2}>0,\
\sgn d\s_{2}=\sgn Q_{10}'$.
From (\ref{ds12}) we also have $\sgn d\s_{1}=-\sgn Q_{20}'$.
These relations define {\it global} orientation
for each connective component in equal time slice,
shown by arrows on solid lines on \fref{f5b}.
Arrows on dashed lines define the direction into the future.
These two kinds of arrows are everywhere related by the rule,
depicted on \fref{fA21}. 

Finally, from (\ref{dp0dl1}) we have (\ref{dp0dl}).

\section*{Appendix 3. Geometrical interpretation of Regge condition}

\noindent
In center-of-mass frame $P_{\mu}=(\sqrt{P^{2}},\vec 0)$ we have
$M_{ij}=\half\oint Q_{i}dQ_{j}=\half S_{ij}$,
$S_{ij}$ is oriented area, restricted by a projection of supporting curve 
onto a plane $(ij)$. For this area the following inequalities are valid:
$|S_{ij}|\leq L_{ij}^{2}/4\pi\leq L^{2}/4\pi=P^{2}/\pi$,
where $L_{ij}$ and $L$ are total lengths of supporting curve
in projection to $(ij)$-plane and to space component of CMF respectively.

\vspace{2mm}
\noindent\underline{{\it Remarks.}}

\vsp\noindent
1. Here the first inequality follows from well-known geometrical fact:
among all planar curves the ratio (~area~/~square of length) reaches
maximum on a circle. The last identity is due to $2P_{0}=\int_{0}^{2\pi}d\s
a_{0}=\int_{0}^{2\pi}d\s|\vec Q'|=L.$

\vsp\noindent
2. Equality $|M_{ij}|=P^{2}/2\pi$
is reached on the supporting curve, whose projection to space component
of CMF is a circle (lying in $(ij)$-plane). Correspondent string is a straight
line, rotating at constant angular velocity in CMF \cite{straight}.

\vsp\noindent
3. For exotic solutions this argumentation is not valid.
In this case $M_{ij}=S_{ij}/2$ as earlier, but $2P_{0}=L_{+}-L_{-}$,
where $L_{+}$ is total length of parts of supporting curve with $a_{0}(\s)>0$; 
$L_{-}$ is length of parts with $a_{0}(\s)<0$.
Generally $2P_{0}\leq L=L_{+}+L_{-}$.

\section*{Appendix 4. DDF variables and light cone gauge}

\noindent
1. Normal solutions.
For given set $\{\vec A_{n}\}$ the function $\vec Q'_{\perp}(\s)$
can be reconstructed: $\vec Q'_{\perp}(\s)={\textstyle{{1}\over{2\pi}}}
\sum_{n}\vec A_{n}e^{-in\s}$, then using the relation
$Q_{+}'={\textstyle{{\pi}\over{2P_{-}}}}\;\vec Q_{\perp}'^{2}$
and one integration, we can find the supporting curve $Q_{\mu}(\s)$.

\begin{center}
\begin{figure}\label{f9}
\parbox{6cm}{~\epsfysize=3.5cm\epsfxsize=6cm\epsffile{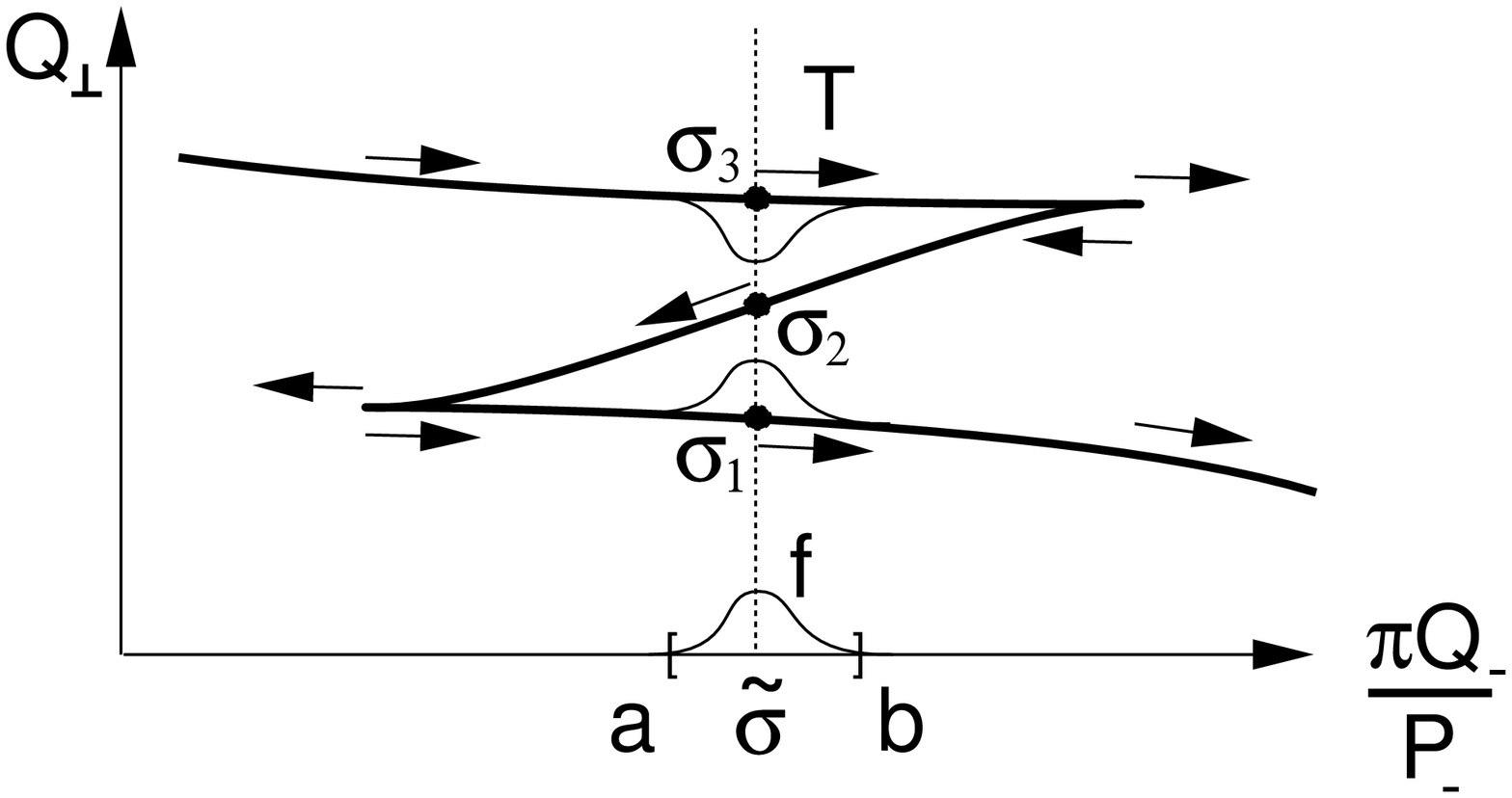}}
\quad\parbox{6cm}{
\fignum DDF variables are defined by sum of vectors $T$
in three points $\s_{1,2,3}$.}
\end{figure}
\end{center}

\vsp\noindent
2. Exotic solutions.
In this case we cannot reconstruct the function
$\vec Q'_{\perp}(\s)$: $${\textstyle{{1}\over{2\pi}}}\sum_{n}
\vec A_{n}e^{-in\tilde\s}=\int_{0}^{2\pi}d\s\;\vec Q'_{\perp}(\s)\;
\Delta\left(\pi Q_{-}(\s)/P_{-}-\tilde\s\right)={\textstyle{{|P_{-}|}\over{\pi}}}
\sum_{i}\vec T_{\perp}(\s_{i}),$$ here $\Delta(\s)={\textstyle{{1}\over{2\pi}}}
\sum_{n}e^{in\s}$ is $2\pi$-periodical delta-function;
$\s_{i}$ are solutions of the equation\\ $\pi Q_{-}(\s)/P_{-}=\tilde\s$;
$\vec T_{\perp}=\vec Q_{\perp}'/|\vec Q_{-}'|$ is transverse component
of tangent vector $T_{\mu}=Q_{\mu}'/|Q_{-}'|$, see \fref{f9}.
For non-monotonous intervals of supporting curve we can reconstruct
not vectors $\vec T_{\perp}$ themselves (they are sufficient to obtain
the supporting curve), but {\it their sum} in a slice $Q_{-}=Const$.
Obviously, we can deform the supporting curve in such a way,
that this sum will not be 
changed. Particularly, we can add to $\vec T_{\perp}(\s)$
near point $\s_{1}$ any function of the form $\delta\vec T_{\perp}(\s)=
\vec f(\pi Q_{-}(\s)/P_{-})$, where $\vec f\in C^{\infty}$ and has finite
support $[ab]$, and subtract this function
in the vicinity of point $\s_{3}$. Thus, for exotic solutions
variables $\{\vec A_{n}\}$ do not define the shape of supporting curve
uniquely. 

\section*{Appendix 5. Tangent angle gauge}

\def\v{\wedge}
\def\P{{\cal P}}
\def\Z{{\bf Z}}
\def\ratio{\frac}
\def\rat#1#2{{\textstyle\ratio{#1}{#2}}}

Let's transfer the mechanics to CMF: introduce orthonormal basis
of vectors $N_{\mu}^{\alpha}(P):\ N_{\mu}^{\alpha}N_{\mu}^{\beta}=g^{\alpha\beta}$,
with $N_{\mu}^{0}=P_{\mu}/\sqrt{P^{2}}$, and decompose the supporting curve
by this basis: $Q_{\mu}(\s)=N_{\mu}^{\alpha}Q^{\alpha}(\s),\ 
a_{\mu}(\s)=N_{\mu}^{\alpha}a^{\alpha}(\s),$ etc. For the components $a^{\alpha}(\s)$
we write the following parametrization:
\begin{equation}
a^{\alpha}(\sigma)=a^{0}(\sigma)\ (1,\cos(\sigma+\theta),\sin(\sigma+\theta)),
\quad a^{0}(\s)=\sum\nolimits_{n}\alpha_{n}e^{in\s},\ 
\alpha_{n}^{*}=\alpha_{-n}.\label{plpar}
\end{equation}
From the condition $\int_{0}^{2\pi}d\s a^{\alpha}(\s)=2(\sqrt{P^{2}},\vec 0)$
we have $\alpha^{0}= \sqrt{P^{2}}/\pi,\ \alpha_{1}=\alpha_{-1}=0.$

Substituting this parametrization into general symplectic form,
defining the string dynamics\footnote{For the details of this technique
see \cite{straight,ax}.}, obtain:
\begin{eqnarray}
&&\Omega=dP_{\mu}\v dZ_{\mu} - dS\v d\theta 
-i\pi\sum_{n\geq2}\ratio{1}{n(n^{2}-1)}\;d\alpha_{n}\v
d\alpha_{n}^{*},\quad Z_{\mu}=X_{\mu}+S\Gamma_{\mu},\nn\\
&&X_{\mu}=Q_{\mu}(0)+\int_{0}^{2\pi}d\s{{a^{0}q_{\mu}}\over{2\sqrt{P^{2}}}}
+{{P_{\mu}}\over{2\pi\sqrt{P^{2}}}}\int_{0}^{2\pi}d\s q^{0},\
q^{\alpha}(\s)=\int_{0}^{\s}d\tilde\s a^{\alpha}(\tilde\s),\
\Gamma_{\mu}=N_{\nu}^{1}\df N_{\nu}^{2}/\df P_{\mu}.\nn 
\end{eqnarray}
$\bullet$ Here $Z_{\mu}$ is canonical mean coordinate,
conjugated to $P_{\mu}$ (compare with \cite{straight}).
From these expressions one can obtain $Q_{\mu}(0)$ as a function
of $Z_{\mu}$ and oscillator variables, and then reconstruct the supporting curve as
$Q_{\mu}(\s)=Q_{\mu}(0)+q_{\mu}(\s)$. 

\noindent
$\bullet\ S=\half\oint Q^{1}dQ^{2}$ is string's angular moment in CMF 
(spin). Here we consider $d=(2+1)$-dimensional Minkowsky space, where 
the spin is a scalar variable, which can take both signs. 

\vsp
The obtained symplectic form corresponds to the following Poisson brackets:
$$\{Z_{\mu},P_{\nu}\}=g_{\mu\nu},\quad \{S,\theta\}=1,\quad
\{\alpha_{n},\alpha_{m}^{*}\}=-i\pi^{-1}n(n^{2}-1)\delta_{nm},\ n,m\geq2.$$
It's convenient to introduce new oscillator variables:
$$a_{n}=\pi^{1/2}(n(n^{2}-1))^{-1/2}\ \alpha_{n},\quad
\{ a_{n},a_{m}^{*} \}=-i \delta_{nm}.$$

Substituting (\ref{plpar}) into the definition of spin, we obtain 
a constraint (mass shell condition):
\begin{equation}
\Phi={{P^{2}}\over{2\pi}}-S-\sum_{n\geq2}n\;a_{n}^{*}a_{n}=0.
\label{constr}
\end{equation} 
This constraint should be used as Hamiltonian of the system.
One can easily prove, that it generates the reparametrization of supporting curve  
$Q_{\mu}(\s)\to Q_{\mu}(\s+\tau)$.

Generators of Lorentz group are defined by expression, analogous to written
in \cite{straight}:
\begin{eqnarray}
&&M_{\mu\nu}=X_{[\mu}P_{\nu]}+SN_{[\mu}^{1}N_{\nu]}^{2}.\label{Lor}
\end{eqnarray}
They generate Lorentz transformation of a coordinate frame
$(N_{\mu}^{0},N_{\mu}^{k}e^{k}_{i}),\ \vec e_{1}=(\cos\theta,\sin\theta),\
\vec e_{2}=(-\sin\theta,\cos\theta)$, by which the supporting curve is decomposed
with scalar ($\theta$-independent) coefficients. 

\vspace{3mm}
\noindent\underline{{\it Remarks.}}

\begin{figure}\label{fA30}
\parbox[c]{7cm}{
~\epsfysize=2.3cm\epsfxsize=7cm\epsffile{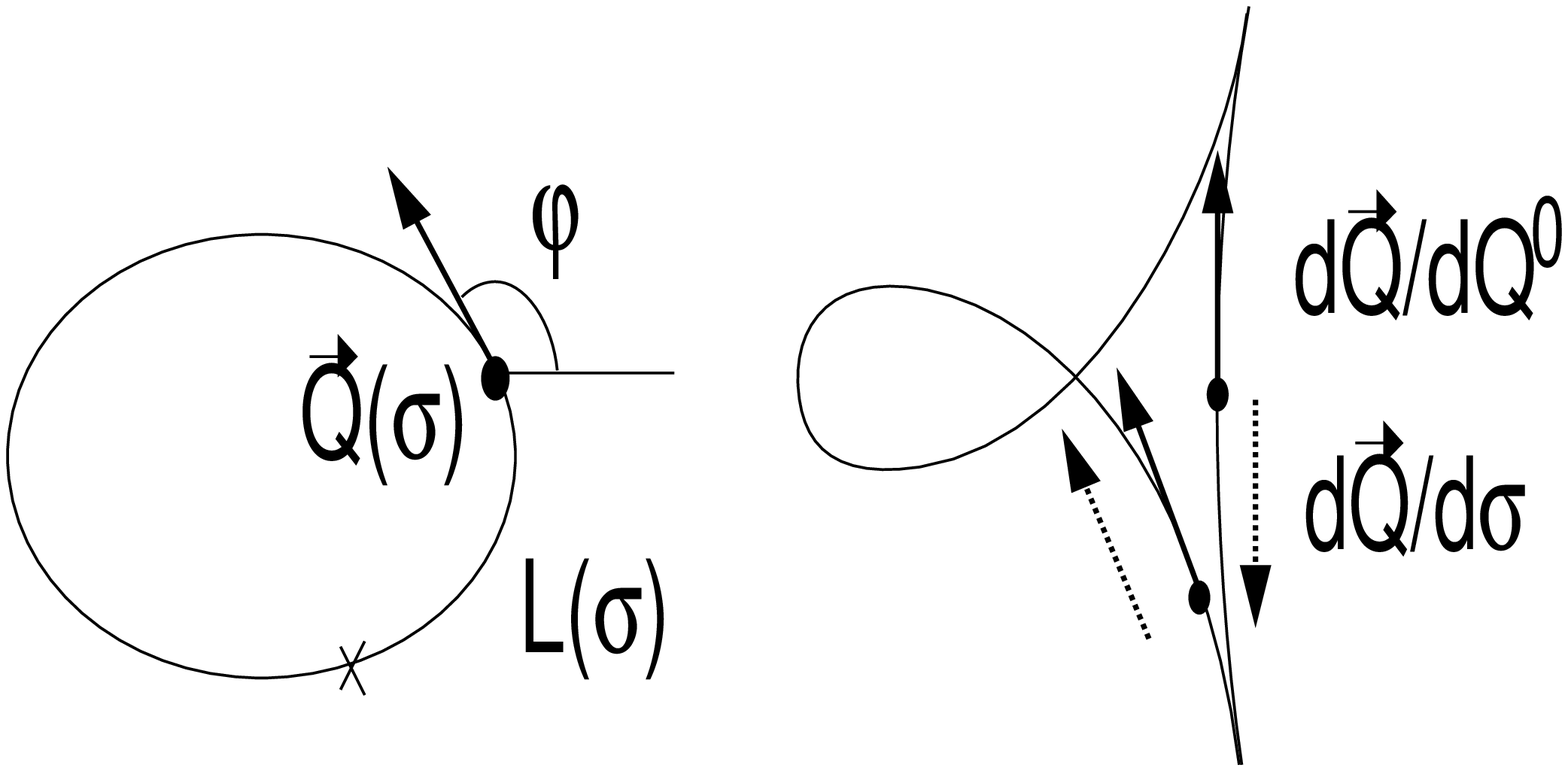}
}\quad\parbox[c]{5cm}{
\fignum Definition of TAG. For exotic solutions (right part)
on the intervals of supporting curve, directed into the past,
$d\vec Q/d\s=a^{0}(\s)d\vec Q/dQ^{0},\ 
d\vec Q/dQ^{0}=(\cos(\s+\theta),\sin(\s+\theta)),\ 
d\vec Q/d\s\uparrow\downarrow d\vec Q/dQ^{0}$.
Polar angle of vector $d\vec Q/dQ^{0}$ is monotonic in $\s$.
}
\end{figure}

\vsp\noindent 1. Parameter $\ph=\s+\theta$ is equal to polar
angle of vector $\vec Q'(\s)$, tangent to the supporting curve on CMF plane, 
see 
\fref{fA30}. 
(That's why we call this parametrization as {\it tangent angle gauge, TAG}). 
Function $a^{0}(\sigma)$ is equal to the radius of curvature 
$a^{0}=dQ^{0}/d\s=dL/d\ph=R$.

\vsp\noindent 2. TAG describes only those supporting curves,
for which $(i)$ during the passage along the curve the tangent vector is 
rotated 
always in the same direction (supporting curve has no inflection points),
and $(ii)$ this vector performs one complete revolution in one complete passage
(winding number of the curve $\nu=1$).
According to \cite{sing}, such supporting curves correspond to a class
of initial data, for which {\it singular points} do not appear in the
evolution. (Generally, the dynamics of open string in $(2+1)$-dimensional
Minkowsky space contains stable singularities, which have a form
of cusps, moving along the string at light velocity.
Number of {\it permanent} cusps, which do not appear or disappear,
equals $\nu-1$. Inflections of the supporting curve correspond to
creation/annihilation of cusps. For the supporting curves
with $\nu=1$ and without inflections the cusps are absent.)
TAG also includes the supporting curves, for which the function 
$a^{0}(\sigma)$ 
is not positive (those, for which $P^{2}>0$ and polar angle of vector 
$d\vec Q/dQ^{0}$ is monotonic in $\s$, see \fref{fA30}).
Thus, TAG includes the solutions from normal sector, for which the string 
has no
cusps, and a definite admixture of solutions from exotic sector.

\begin{figure}\label{fA31}
\parbox[c]{7cm}
{~\epsfysize=4.5cm\epsfxsize=3cm\epsffile{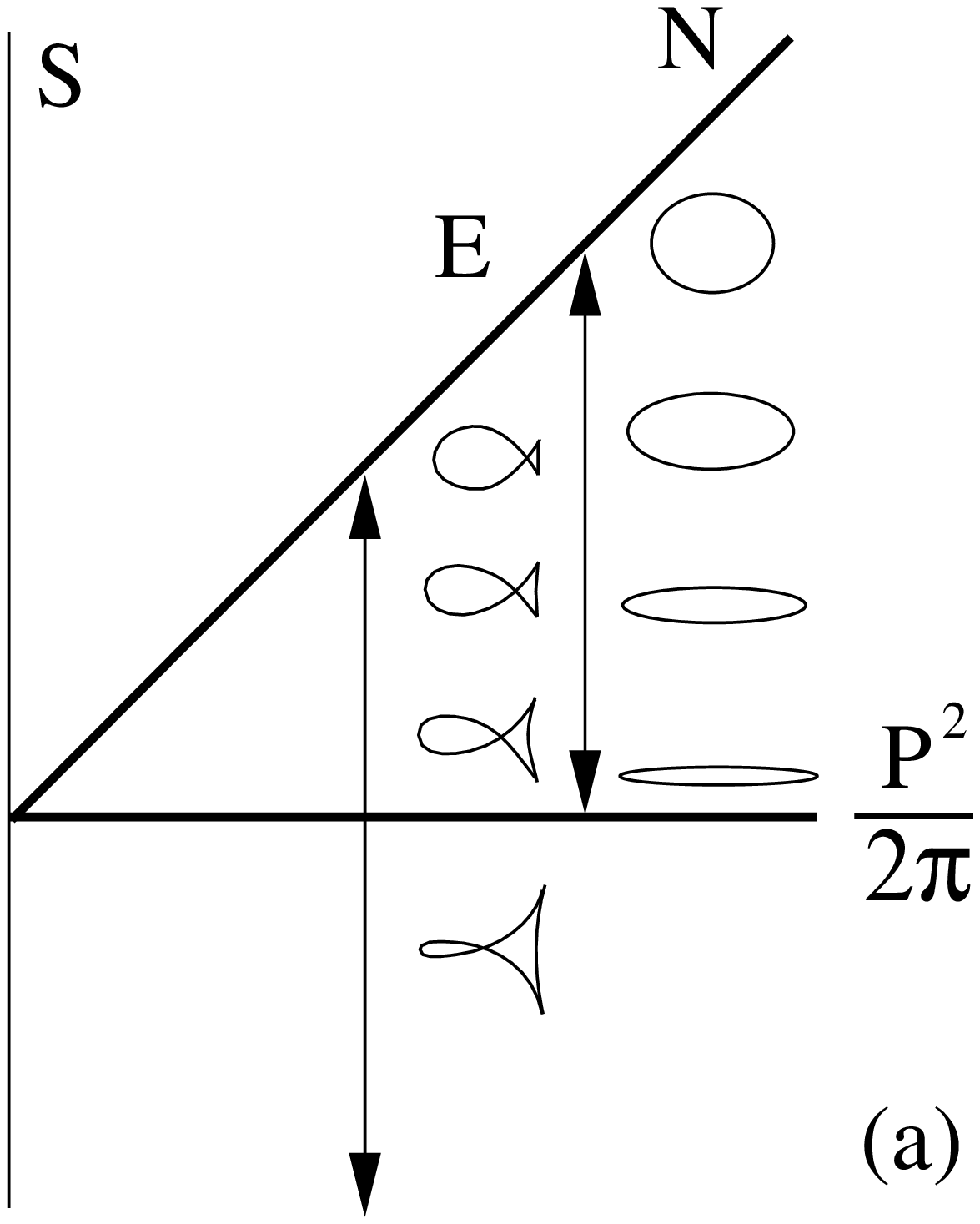}
\quad~\epsfysize=4.5cm\epsfxsize=3cm\epsffile{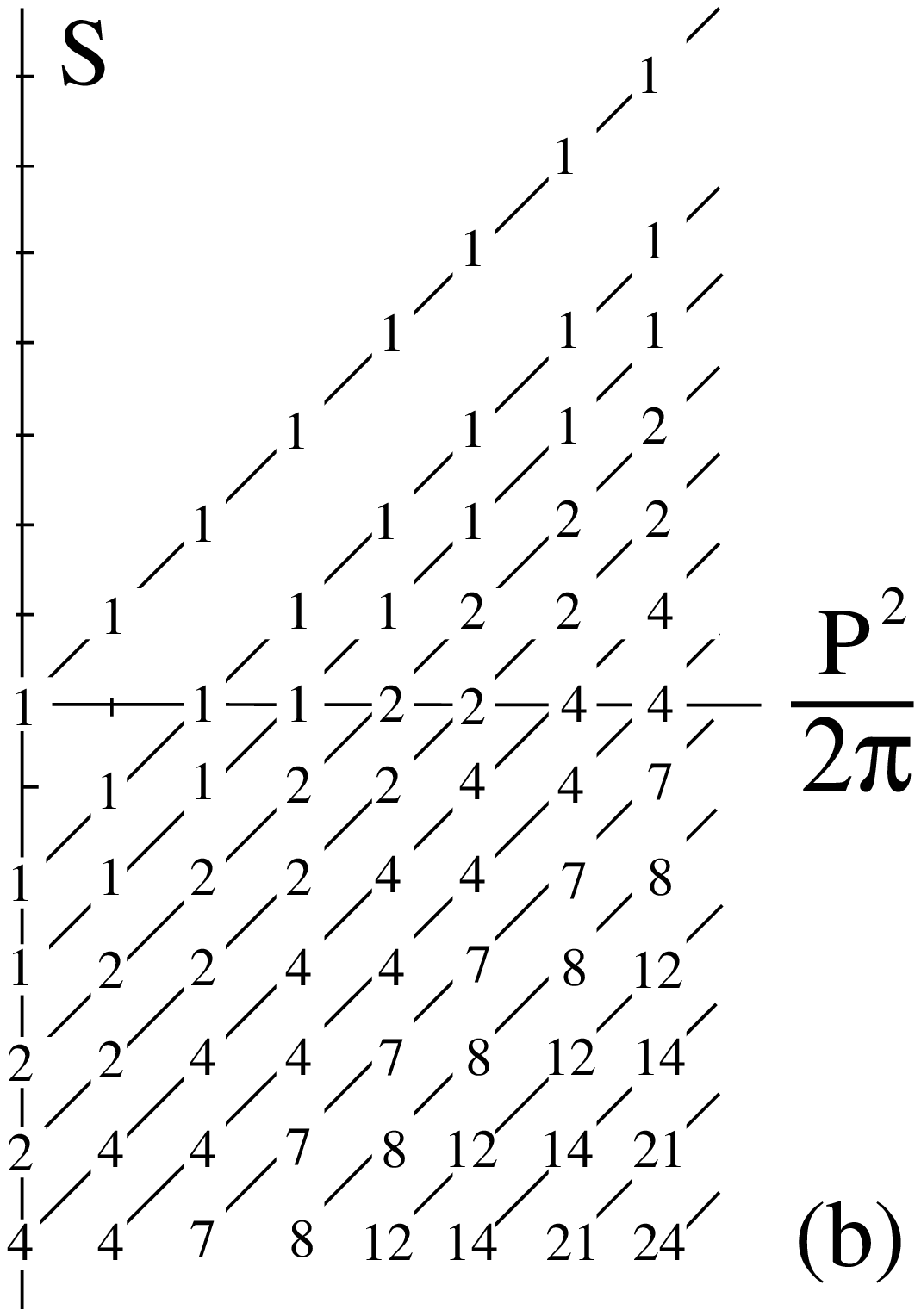}
}\quad\parbox[c]{5cm}
{\fignum a) Region of variables $(P^{2}/2\pi,S)$ in TAG. 
E -- exotic solutions, N -- normal solutions. 
b) Quantum spin-mass spectrum of TAG.
}
\end{figure}

\vsp\noindent 3. Region of variation of variables $(P^{2}/2\pi,S)$ and typical
examples of supporting curves are shown on \fref{fA31}a. 
Normal solutions occupy on this plane the sector $0\leq S\leq P^{2}/2\pi$.
For exotic solutions $S\leq P^{2}/2\pi$, here $S$ is not bounded from below 
(see (\ref{constr})), exotic solutions violate the Regge condition. 

\vspace{2mm}
\noindent

\noindent

\paragraph*{Quantization} of this mechanics is straightforward.
Canonical operators:
\begin{eqnarray}
&&[Z_{\mu},P_{\nu}]=ig_{\mu\nu}\quad [S,\theta]=i,\quad
[a_{n},a_{m}^{+}]=\delta_{nm},\ n,m\geq2.\nn
\end{eqnarray}
Oscillator part of the mechanics can be realized in a positively defined Fock space:
$a_{k}|0\ran=0,\ |\{n_{k}\}\ran=\prod_{k\geq2}
\frac{1}{\sqrt{n_{k}!}}(a^{+}_{k})^{n_{k}}|0\ran,$
with the occupation numbers $n_{k}=a^{+}_{k}a_{k}=0,1,2...$
Complete space of states can be defined as a direct product of this Fock space
onto the space of functions $\Psi(P_{\mu},\theta)$, $2\pi$-periodic for $\theta$, 
with definition of operators $Z_{\mu}=i\df/\df P_{\mu}$, $S=i\df/\df\theta$ 
(eigenvalues $S\in\Z$). 
Physical subspace is selected by the constraint
$({{P^{2}}/{2\pi}}- S-\sum kn_{k})|phys\ran=0.$ 
Correspondent spin-mass spectrum is shown
on \fref{fA31}b. It consists of linear Regge trajectories
$P^{2}/2\pi-S=T,\ T=\sum_{k\geq2}kn_{k}=0,2,3...$.
The multiplicity of states is constant on the trajectory 
and is equal to multiplicity of eigenstates of operator $T$, 
e.g. $T=0:\ |0\ran,\ T=2: |1_{2}\ran,\ T=3: |1_{3}\ran,\ T=4: |2_{2}\ran,|1_{4}\ran$
(2 states), etc.

\vspace{2mm}
\noindent\underline{{\it Remark:}}
in our approach $Z_{\mu},P_{\mu}$ and $S$ are taken as {\it independent} 
canonical variables, for which the commutators are postulated directly from 
Poisson 
brackets. It was shown in \cite{straight}, that in this case the Lorentz 
generators,
defined by expression (\ref{Lor}), have no anomaly in commutators.

% \baselineskip=\normalbaselineskip\normalsize


\begin{thebibliography}{99} 
   \bibitem{zone}
      \BY{G.P. Pronko}
      \IN{Rev.Math.Phys.}{2}{1990}{355}.
   \bibitem{Rohrlich}
      \BY{F. Rohrlich}
      \IN{Phys.Rev.Lett.}{34}{1975}{842}.
   \bibitem{DDF}
      \BY{E. Del Giudice, P. Di Vecchia \atque A. Fubini}
      \IN{Ann.Phys.}{70}{1972}{378}.
   \bibitem{Blokhincev}
      \BY{V.N. Pervushin \atque A.M. Khvedelidze}
      ``Gaugeless Quantization of Constrained Systems'',
      pp.104-132 in Proc. of seminars in memory of D.I.Blokhintsev, 
      published by JINR, Dubna 1995.
   \bibitem{Zhelt}
      \BY{A.A. Zheltuhin}
      \IN{Sov.J.Nucl.Phys.}{34}{1981}{562}.
   \bibitem{Brink}
      \BY{L. Brink \atque M. Hennaux}
      \TITLE{Principles of String Theory},
      Plenum Press, New York and London 1988.
   \bibitem{Groebner}
     \BY{B. Buchberger} ``Groebner bases: An algorithmic method 
     in polynomial ideal theory'', in the book 
     \TITLE{Multidimensional systems theory}, ed. N.K.Bose, 
   Dordrecht: Reidel, 1985, p.207. 
   \bibitem{DDF_theorem}
      \BY{R.C. Brouwer}\IN{Phys.Rev.}{D6}{1972}{1665}; 
      \BY{P. Goddard \atque C.B. Thorn}\IN{Phys.Lett.}{40B}{1972}{235}. 
   \bibitem{indef}
      \BY{I.N. Nikitin}
      \IN{Theor.Math.Phys.}{107}{1996}{589}.
   \bibitem{straight}
      \BY{G.P.Pronko \atque A.V. Razumov}
      \IN{Theor.Math.Phys.}{56}{1983}{192}.
   \bibitem{par6}
      \BY{I.N. Nikitin}
      \IN{Theor.Math.Phys.}{109}{1996}{1400}.
   \bibitem{ax}
      \BY{I.N. Nikitin}
      \IN{Sov.J.Nucl.Phys.}{56}{1993}{230}.
   \bibitem{vis}
      \BY{S.V. Klimenko, V.V. Dyachin \atque I.N. Nikitin}
``Singularities on the world sheets of open relativistic strings''
chap.18 in the book \TITLE{Scientific Visualization: Overviews, Methodologies,
and Techniques}, IEEE Comp.Society Press, Los Alamitos 1997.
   \bibitem{sing}
      \BY{S.V. Klimenko \atque I.N. Nikitin}
      \IN{Theor.Math.Phys.}{114}{1998}{299}.
 \end{thebibliography}
\end{document}